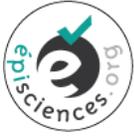 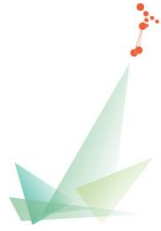 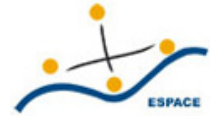

# Réseau politique des agents du pouvoir central : l'exemple des *missi dominici*[1]

**Andrey Grunin**

CIHAM - UMR 5648, Université de Lyon 2, Avignon Université, France

Correspondance : grunin.andrey@gmail.com

**Résumé**

Cette étude propose plusieurs modèles d'analyse d'un réseau politique des agents du pouvoir central du haut Moyen Âge, les *missi dominici*. Complétées par l'analyse statistique, plusieurs hypothèses de recherche fondées sur les positions de l'historiographie actuelle ont été étayées. D'une part, l'examen du réseau des agents a permis de mettre en lumière une certaine évolution liée à la structuration du système missatique et au mécanisme de la transition des agents d'un règne à l'autre. D'autre part, l'étude des relations entre les agents et les lieux de leurs missions a confirmé une certaine amplification, avec le temps, de la politique du recrutement des *missi* au sein de l'aristocratie locale. Enfin, plusieurs difficultés liées tant au caractère lacunaire des données issues des sources médiévales qu'à la complexité de modélisation et d'analyse d'un réseau politique multimodal ont été évoquées.

**Mots-clés**

Réseaux politiques ; Analyse de réseaux ; Graphes bipartis ; Réseaux d'affiliations ; *Missi dominici*

---

[1] Les premiers jalons de ce travail ont été présentés lors du séminaire de recherche de l'équipe SAMM (Statistique, Analyse et Modélisation Multidisciplinaire) (EA 4543, Université Paris 1 Panthéon-Sorbonne) organisé par Julien Randon-Furling le 16 juin 2017.

# I  INTRODUCTION

L'objet de cette recherche est un réseau d'agents du pouvoir central (*missi dominici*) mis en place par les Carolingiens en Europe occidentale durant le premier Moyen Âge. Bien que, comme le montre Hanning (1983), la pratique d'appel aux *missi* semble déjà être apparue vers la moitié du VIII$^e$ siècle, c'est le règne de Charlemagne qui a marqué un tournant important : le recours à ces agents est devenu régulier et le système a survécu, peu ou prou, jusqu'à la fin du IX$^e$ siècle. La première mission répertoriée pour cette recherche date de 751 et la dernière de 886. Plusieurs capitulaires ont fixé les lignes directrices de l'exercice du pouvoir des *missi dominici* et ont témoigné de l'évolution de l'institution dans le temps[2]. Si la tâche principale de ces agents a été le contrôle et l'inspection des territoires sous la domination franque, leurs activités sur place pouvaient être variées, du règlement des conflits juridiques et la diffusion des décrets royaux jusqu'à l'inspection des marchés (Ganshof 1965 ; Werner 1980 ; Davis 2015). Constitués généralement d'un laïc et d'un ecclésiastique, ces groupes d'agents royaux devaient effectuer des tournées régulières dans les différentes parties du royaume regroupées en zones d'inspection, *missatica*. Cependant le contour aussi bien géographique que temporel des *missatica* suscite plusieurs interrogations. Si quelques textes les ont mentionnés de façon explicite, les autres n'ont donné que des informations lacunaires sur les lieux d'exercice des *missi dominici*. Le travail de reconstitution de ces zones d'inspection, déjà commencé par Eckhardt (1956), mérite d'être poursuivi et approfondi. Hors des *missi* « ordinaires » inscrits dans le cadre de leur circonscription, on peut faire état des *missi* « extraordinaires » envoyés pour un objectif ponctuel dans les territoires périphériques (Ganshof 1965 ; Werner 1980 ; Bougard 1995) ou des *missi* en tant qu'ambassadeurs (Scior 2009 ; Kikuchi 2013). Enfin, la fréquence des tournées pouvait varier au fil des années et en fonction des terres contrôlées.

Une des premières balises dans l'étude des *missi dominici* a été posée par le travail de Krause (1890). L'auteur a conclu à l'essor du système missatique exclusivement sous Charlemagne et a constaté que les agents ont été, dans la plupart des cas, recrutés parmi les *vassi dominici* (vassaux royaux) extérieurs aux lieux de leurs missions. Ces constats ont marqué en partie l'historiographie du début de XX$^e$ siècle, comme en témoigne par exemple le travail de Thompson (1903). Durant les décennies suivantes, le sujet a connu un intérêt modéré parmi les médiévistes. Les quelques travaux qui en résultèrent commençaient à nuancer les positions historiographiques précédentes. La vision étroite de l'essor de l'institution des *missi* sous le règne de Charlemagne a été mise en doute par des enquêtes poussées démontrant l'évolution progressive du système tout au long du IX$^e$ siècle (Eckhardt 1956 ; Ganshof 1965 ; Werner 1980). La disparition du réseau des *missi* semble être davantage causée par les transformations profondes, sociales et économiques, de la royauté franque que par la faiblesse de l'institution missatique elle-même. En même temps, les nouvelles études ont affirmé que l'appel aux *vassi dominici*, souvent enclins à la corruption, a été rapidement abandonné au profit du recrutement des *missi* parmi les membres des familles aristocratiques locales ou parmi les élites ayant déjà un ancrage fort dans les territoires de leurs affectations (Eckhardt 1956 ; Hanning 1984b). Plus récemment, le travail prosopographique de Kikuchi (2013) a marqué une étape majeure dans l'étude du système missatique. Grâce au recours à une riche palette de sources, il a tenté de reconstituer les activités et le parcours du plus grand nombre de *missi* connus à ce jour et a montré les changements dans l'utilisation de ce réseau des agents au cours des différents règnes. Jusqu'à ce jour, les études de l'ensemble du système missatique sont toutefois rares et il n'est possible de recenser qu'une dizaine de travaux sur le sujet (à l'exception de ceux déjà cités : de Clercq 1968 ; Hanning 1983 ; Hanning 1984a ; Depreux 2001 ; Gravel 2007 ; McKitterick 2009 ; Jégou 2010).

---

[2] *Duplex legationis edictum* (789), Capit. I, pp.62-64, *Capitulare missorum generale* (802), Capit. I, pp.91–99 et *Capitularia missorum specialia* (802), Capit. I, pp.99–104. Sur le dernier texte voir Eckhardt (1956). Sans oublier les écrits postérieurs : *Commemoratio missis data* (825), Capit. I, pp.308-309 ; *Capitulare Missorum Silvacense* (853), Capit. I, pp.270-276.

Finalement, le caractère parfois incertain aussi bien des fonctions que de la régularité et de l'ampleur des pratiques missatiques pouvait laisser subsister quelques interrogations sur la véritable nature institutionnelle du recours à ces agents par la cour royale franque. L'étude de ce réseau s'inscrit alors dans le cadre d'une recherche plus large sur le fonctionnement du système politique altomédiéval et sur les dynamiques de gouvernance aux temps pré-modernes. Si les questions sur les formes et l'existence même de l'État médiéval restent toujours ouvertes (Werner 1992 ; Reynolds 1997 ; Davies 2003 ; Grunin 2019) un des points de vue dominants de l'historiographie propose de le percevoir comme un réseau complexe d'influence et de domination (Werner 1980 ; Althoff 2004 ; Dumézil 2013). Dans cette optique, l'étude des agents royaux doit nous permettre d'observer au plus près la nature de cette structure du pouvoir.

Les points de départ de cette enquête se sont fondés ainsi sur deux postulats de l'historiographie actuelle :

i) Bien que le réseau des *missi dominici* semble avoir existé tout au long du IX$^e$ siècle, son essor a été marqué par les règnes de trois souverains : Charlemagne, Louis le Pieux et Charles le Chauve. La survivance du système des agents royaux durant plusieurs décennies témoigne de l'existence d'une réelle structure politique et pas seulement d'une spécificité de gouvernance d'un seul souverain. La première étape de cette recherche consiste alors à étudier les caractéristiques propres du réseau des agents du pouvoir central et à mettre en évidence les changements qui s'y opéraient au fil du temps.

ii) La position historiographique actuelle fait état du recrutement, dans la majorité des cas, des *missi* en raison de leur appartenance aux familles liées aux pouvoirs en place ou à l'élite qui y exerçait déjà des fonctions. L'étude plus détaillée des liens qui reliaient les agents aux lieux de leurs affectations vise à examiner de plus près ce constat. La présence des relations significatives attestera la volonté de mettre en place un mécanisme de contrôle des territoires basé sur les forces locales.

L'ensemble de cette étude s'est appuyé sur plusieurs conventions terminologiques. La notion de *fonction* appliquée à l'aristocratie du haut Moyen Âge peut avoir des lectures multiples (Depreux *et al*. 2007 ; Bougard *et al*. 2013). Dans ce travail, ce terme a été utilisé dans le cadre strictement analytique pour désigner la détention des différentes dignités telles que de comte, duc, abbé, etc. De ce fait, par les *lieux de fonctions* on entend les endroits et les possessions foncières auxquelles ces dignités pouvaient être rattachées. Les *lieux d'affectations* des *missi* comprennent uniquement les endroits où l'activité missatique a eu lieu. Les *lieux d'attaches personnelles*, quant à eux, renvoient aussi bien aux lieux d'origines géographiques des *missi* qu'aux endroits où les *missi* pouvaient avoir des liens de parenté.

## II  MÉTHODOLOGIE ET DONNÉES

### 2.1  Méthodologie

Les travaux précédents ont déjà dessiné plusieurs traits du système missatique et ont fourni un nombre important de renseignements. Les données recueillies ont donné la possibilité d'embrasser l'histoire des envoyés des rois carolingiens dans toute son ampleur. Cela a signifié de déplacer l'accent de l'examen des cas uniques à l'exploration des structures institutionnelles à une échelle plus large, aussi bien chronologique que géographique. Ce changement de perspective a nécessité le renouveau du cadre méthodologique. Il fallait aussi bien tenir compte de la complexité des liens qui pouvaient relier les agents entre eux ou avec leurs lieux d'affectations que déceler les transformations qui s'y produisaient. L'analyse de réseaux, centrée sur l'étude des relations et appuyée par l'analyse statistique multivariée répondait parfaitement à ces exigences. Le recours à ces approches en sciences humaines et sociales a déjà une longue histoire et leur description détaillée, largement faite ailleurs, dépasse de loin le propos de ce travail. On renvoie donc à Scott (2000) et Kadushin (2012) qui ont donné une présentation des principes fondamentaux de l'analyse de réseaux, et à Wasserman et Faust (1994), et à Newman (2010) pour leurs argumentations mathématiques. L'analyse statistique et le traitement des données longitudinales ont été de même amplement abordés aussi bien de façon générale (à titre d'exemple : Crawley 2015 ; Hamilton 1994) que dans le contexte de leur utilisation en histoire (Hudson et Ishizu 2016 ; Lemercier et Zalc 2019 et plus généralement pour les sciences sociales : Kolenikov *et al*. 2010).

Il n'est pas inutile toutefois d'attirer attention sur quelques spécificités, tantôt réelles, tantôt apparentes, que l'utilisation de ces techniques, et notamment celle de l'analyse de réseaux, peut révéler (Lemercier 2005 ; Collar *et al*. 2014 ; Brughmans *et al*. 2016). D'une part, une distinction est requise entre la notion de *réseau* introduite sous l'effet de mode dans les enquêtes historiques et dont l'utilisation dépasse rarement le sens métaphorique du terme, et l'*analyse de réseaux* en tant qu'un ensemble défini des techniques avec son propre vocabulaire et sa méthodologie (Lemercier 2005 ; Collar *et al*. 2014 et dans une moindre mesure Collar 2015). Si la lecture symbolique reste encore dominante en sciences humaines, cela est dû, comme le souligne à juste titre Beauguitte (2016), à la complexité supposée de la dimension technique de l'approche. D'autre part, l'incomplétude des informations issues des sources lacunaires semble entraver le déploiement de la méthode qui exige des renseignements précis, obstacle toutefois surmontable et déjà surmonté au sein des autres disciplines dans lesquelles la quête des données exhaustives a été depuis longtemps abandonnée au profit d'un échantillonnage (Thompson 2012). En dépit de ces difficultés, l'analyse de réseau, aidée par la démocratisation des outils informatiques, a progressivement fait son chemin en histoire et notamment en histoire médiévale (Rosé 2011 ; Bouveyron *et al*. 2014 ; Hammond *et al*. 2017).

### 2.2 Corpus de données

Les travaux qui ont servi de base du corpus sont l'étude prosopographique de Kikuchi (2013) et, en second lieu, l'enquête de Krause (1890). Ce dernier a répertorié 214 missions couvrant la période de la deuxième moitié du VIII[e] siècle jusqu'au début du X[e] siècle. Les données retenues sont souvent lacunaires, mais permettent de dessiner les traits majeurs dans la reconstruction des territoires couverts par les *missi* et de retenir quelques informations sur les objectifs de leurs missions. Contrairement à son prédécesseur, Kikuchi (2013) prend les données prosopographiques comme la base de sa recherche. Le dépouillement embrasse les années 751-888 et fournit des informations sur plus de 400 *missi dominici*. Hormis la trajectoire personnelle de chaque agent, cette étude donne des indications relatives à leurs activités en tant qu'envoyés royaux. Les données de ces deux études ont été complétées par des informations ponctuelles issues des différentes enquêtes historiques (Hanning 1983 ; Hanning 1984a ; Hanning 1984b ; Kaiser 1986 ; Poly 1976 ; McCormick 2001 ; McCormick 2011). En outre, il faut citer le travail prosopographique de Depreux (1997) qui donne des renseignements sur 280 individus du règne de Louis le Pieux, dont une partie est relative aux *missi dominici*, et la base numérique « *The Making of Charlemagne's Europe (768-814)* » de Rio *et al*. (2014) qui contient des données sur 64 *missi* du règne de Charlemagne. La reconstitution des liens entre les souverains, les agents et les lieux de leurs missions a également incité à consulter les travaux prosopographiques d'un cadre temporel plus large, notamment les études d'Ebling (1974) et de Borgolte (1986). Cependant il était nécessaire de se limiter dans l'étendue de certaines données, notamment généalogiques. Si la reconstitution de ce type de liens devait permettre de mieux cerner une des caractéristiques du réseau étudié, la recherche plus approfondie des relations familiales au sein de l'aristocratie franque dépassait largement le cadre de ce travail.

Toutes les données utilisées dans cette recherche sont issues des enquêtes historiques précédemment citées. La consultation des sources a été ponctuelle en cas de doutes sur des informations précises. Pour la présentation détaillée des corpus mobilisés, on renvoie à Kikuchi (2013) et, en tout dernier lieu, à Krause (1890). Il est toutefois indispensable de s'attarder sur quelques difficultés liées tant à la nature des sources qu'à leur vocabulaire.

La reconstitution de l'histoire missatique s'est appuyée sur une documentation hétérogène, aussi bien normative que narrative. Une grande partie du corpus a été, par exemple, constituée des capitulaires qui sont une des sources majeures pour dresser les lignes directrices des activités des *missi dominici* : missions commandées, noms des agents, lieux d'affectations (de Clercq 1968 ; McKitterick 2009 ; Kikuchi 2012 ; Kikuchi 2013). Or, comme tout texte du haut Moyen Age, certains capitulaires présentent un problème inhérent de datation. On s'est fié aux datations proposées et les plus largement admises (Ganshof 1958, notamment pp.108-120 ; Tessier 1967, pp.273-439 ; Bühler 1986, notamment pp.484-490). La reconstitution des éléments du parcours personnel d'un agent (origines géographiques, liens de parenté, etc.) s'est nécessairement reliée aux sources narratives : textes hagiographiques, correspondances ou annales. Bien qu'il soit nécessaire de garder une certaine réserve à l'égard des informations transmises par ce type de documentation, ces textes nous ont donné des renseignements importants. A titre d'exemple on ne citera que le poème écrit par Théodulf d'Orléans après son voyage vers 798 en tant que *missus dominicus* de Charlemagne dans les territoires méridionaux du royaume (Theodulfi, *Contra Iudices*. Sur cette mission voir Krause 1890, appendice I, n°18, p.259 ; Depreux 1997, notice 262, pp.383-384 ; Kikuchi 2013, pp.615-617 ; McKitterick 2009, p.262). Ce texte déjà bien connu et étudié (Collins 1950 ; Magnou-Nortier 1994 ; Tignolet 2011) nous livre non seulement un témoignage précieux, bien que partial, des dessous de la justice franque mais également les noms des villes visitées et les activités exercées par l'évêque d'Orléans. Même si la prudence a été de mise, ces informations ont pu enrichir les données sur les services missatiques de ce proche du premier empereur franc (Kikuchi 2013, pp.615-617).

Les difficultés liées à la nature hétérogène du corpus ont été accrues par l'ambiguïté, souvent présente, de l'emploi dans les sources du vocable *missi* lui-même. Une distinction a dû être faite entre les *missi* (envoyés) et les *missi dominici* (envoyés du roi). Si le premier de ces termes pouvait désigner un envoyé ou un messager relevant parfois d'une tout autre autorité que celle du roi, ce sont uniquement les mentions des *missi dominici* qui nous renvoient aux agents appartenant à l'institution du pouvoir central. La séparation claire entre ces deux catégories a pu être délicate. Le vocable *dominus* lui-même avait une valeur différente selon le contexte et désignait aussi bien les membres de l'élite altomédiévale (Depreux 2001), que les chefs de la royauté franque (Fried 1982 ; Goetz 1987). Une notice du plaid de règne de Louis le Pieux en est un exemple révélateur (Bernard et Bruel 1876, n°3, p.6). Ce texte, qui présente également un problème de datation, mentionne plusieurs *missi* parmi lesquels ceux envoyés par le roi sont difficiles à discerner. Si Krause (1890, appendice I, n°69, p.264) liste une grande partie des *missi* nommés, Depreux (1997, notice 208, pp.338-339) et Kikuchi (2013, pp.561-563), dont la position a été retenue dans ce travail, sont plus prudents et ne retiennent que le seul comte Ostoric, clairement désigné comme *missus dominicus*. La recension ici de tous les cas semblables est hors de propos ; on renvoie donc aux travaux déjà cités qui ont servi pour la construction de la base de données. Retenons seulement qu'en cas de doute c'est la position historiographique la plus récente qui a été systématiquement adoptée.

Cette ambiguïté du vocabulaire a été renforcée par l'absence dans les sources d'une distinction nette entre les *missi dominici* qui opéraient à l'intérieur des terres carolingiennes et ceux envoyés en tant qu'ambassadeurs à l'étranger (Scior 2009 ; Nelson 2000). Plusieurs termes, tels que *legatus*, *missus* et *nuntius* se sont entremêlés dans les textes de l'époque (Dreillard 2001). Les mentions de la délégation expédiée par Charlemagne à Constantinople sont un parfait exemple pour démontrer cette ambivalence. Si les *Annales regni Francorum* mentionnent pour l'année 802 l'envoi d'une *légation* constituée de Helmgaud (comte de Meaux) et de Jessé (évêque d'Amiens) (*Annales regni Francorum*, a.802, p.117), le même texte nous relate pour l'année 803 le retour des *missi dominici* de Constantinople (*Annales regni Francorum*, a.803, p.118. Sur cette mission voir également Depreux 1997, p.408 ; Kikuchi 2013, p.451). La thèse de Dreillard (2001) qui propose de distinguer, dans le cadre des ambassades carolingiennes, les *missi dominici*, qui transmettent les décisions royales, des *legati*, qui ont le pouvoir de négocier, mérite alors d'être nuancée. Comme pour la mission de 802-803 à Constantinople, les sources pouvaient, semble-t-il, appeler les mêmes agents tantôt *legati*, tantôt *missi*. Dans ce cas, et aussi dans d'autres semblables, le caractère équivoque des textes ne laisse pas de place à une quelconque argumentation solide pour trancher dans tel ou tel sens. Le choix a été ainsi arrêté de retenir Helmgaud et Jessé en tant que *missi dominici*. De même, la datation de certaines missions, l'identification des lieux et des agents employés a pu être un objet de débat. S'il était impossible de prendre en compte et de quantifier l'ambiguïté de ces informations, les données retenues correspondent aux positions historiographiques actuellement admises.

Le problème d'imprécision des informations disponibles a concerné tout particulièrement le traitement des données géographiques. L'aspect souvent sommaire de ce type de renseignements a dicté les méthodes de leur encodage. Adalhelmus, archevêque de Bordeaux, a été, par exemple, commandité en 816 comme *missus*, dans la même ville où il exerçait déjà ses fonctions ecclésiastiques (Conc. 2.1, pp.458-406 ; voir également Kikuchi 2013, p.282 ; Krause 1890, appendice II, n°63, p. 287). Arn, archevêque de Salzbourg, quant à lui, a été envoyé en 802 pour une action missatique en Bavière (on ne cite qu'une des deux missions d'Arn pour l'année 802 : Bitterauf 1905, n°183, p.174 cité d'après Kikuchi 2013, pp.285-286 ; sur la même mission Krause 1890, appendice I, n°31, p.261). Si dans le premier cas, le lieu d'affectation se présentait comme une ville et a dû être encodé en tant qu'un point dans l'espace, dans le deuxième cas, le lieu d'affectation correspondait à une région et a dû, par conséquent, être encodé en tant qu'un polygone.

Dans le cadre de l'analyse des relations entre les lieux d'affectations et les lieux d'attaches personnelles ou de fonctions d'un agent, le traitement et l'analyse de ces données ont posé plusieurs difficultés. Dans la configuration la plus simple, quand les lieux d'affectations et de fonctions, par exemple, sont exactement les mêmes, la relation se présente comme existante. C'est le cas d'Adalhelmus pour qui le lieu de fonction et le lieu d'affectation ont été la même ville, Bordeaux. En revanche, dans les cas d'Arn, pour qui le lieu de fonction a été Salzbourg et le lieu d'affectation la Bavière, l'analyse des relations entre ces deux endroits a été plus délicate. Le calcul de la distance euclidienne la plus courte entre les différents lieux a été choisi comme la technique la plus opportune de traitement de ce type d'information. Deux modèles de mesure ont été alors possibles :

- d'une part, la distance la plus courte entre deux points (deux villes) a été calculée comme la distance minimale sur une surface. De ce fait, la distance entre le lieu d'affectation et le lieu de fonction d'Adalhelmus est égale à 0. Les deux points sont les mêmes, la ville de Bordeaux.

- d'autre part, pour la distance la plus courte entre un polygone et un point (une région et une ville), que le point soit contenu ou non dans le polygone, le calcul s'est effectué à partir du barycentre mathématique du polygone jusqu'au point donné. Dans le cas d'Arn, la distance entre son lieu de fonction et son lieu d'affectation a été égale à 80 kilomètres. C'est la distance entre la ville de Salzbourg, encodée comme un point, et le barycentre mathématique de la région Bavière, encodée comme un polygone.

Les mêmes principes de calcul se sont appliqués lors de l'analyse des distances entre les lieux d'affectations et les lieux d'attaches personnelles. Dans les cas où l'agent avait plusieurs lieux de fonctions (ou d'attache personnelle), la distance a été calculée pour le lieu de fonction (ou d'attaches personnelles) le plus proche de son lieu d'affectation. La nature parfois incertaine des données géographiques incite toutefois à garder des réserves face aux résultats obtenus. Une précaution supplémentaire concerne l'impossibilité d'établir avec certitude le cadre temporel dans l'évolution des lieux de fonctions ou d'attaches personnelles d'un agent. Les informations souvent disparates des sources médiévales n'ont pas permis de corréler les dates des missions avec les dates du parcours personnel de chaque *missus*.

Enfin, le caractère lacunaire des sources médiévales a conduit inévitablement au problème des données manquantes. Pour un grand nombre d'agents ou pour certaines missions, il était impossible de reconstituer l'ensemble des renseignements (Table 1).

|  | **Données connues, nb. (pourcentage)** | **Données inconnues, nb. (pourcentage)** | **Nombre total** |
|---|---|---|---|
| **Missions (lieu)** | 394 (99%) | 4 (1%) | |
| **Missions (date)** | 398 (100%) | — | **398 missions** |
| **Missions (objectif)** | 371 (93%) | 27 (7%) | |
| **Agents (fonction)** (comte, évêque, etc.) | 315 (66%) | 160 (34%) | |
| **Agents (lieu de fonction)** | 211 (44%) | 264 (56%) | **475 agents** |
| **Agents (lien de parenté)** (avec d'autres agents ou appartenance familiale) | 68 (14%) | 407 (86%) | |
| **Souverains*** | — | — | **14 souverains*** |
| **Lieux** | — | — | **307 lieux** |

Table 1. Résumé des données manquantes et connues.

\* - Les dénominations des dirigeants francs des VIIIe-IXe siècles ont connu de nombreuses évolutions ; les titres royaux et impériaux se sont côtoyés dans les titulatures des Carolingiens à partir de l'an 800 (Schneider 1997 ; Sot 2007). Dans le cadre de cette recherche, il a été fait occasionnellement appel au vocable *souverain*, sans aucune référence à la notion de *souveraineté* des siècles postérieurs.

Dans les cas les moins riches en informations, on n'a disposé que des noms des agents, du souverain qui les a envoyés et de la date et du lieu de la mission. L'incomplétude des données a soulevé par conséquent la question du nombre réel des agents existant dans le cadre chronologique étudié. Même si quelques textes normatifs ont exprimé la volonté de mettre en place une structure fixe de l'emploi des *missi dominici* (par exemple *Capitulare missorum generale* 802, Capit. I, pp.91-99 ; *Capitularia missorum specialia* 802, Capit. I, pp.99-104), l'absence de données fiables sur la durée effective du service d'un agent et sur la fréquence de leur rotation a rendu impossible une quelconque estimation de leur nombre exact. L'absence d'une partie importante de données, notamment celles sur les liens entre les *missi*, a posé, dans le cadre de l'analyse de réseaux, des difficultés dans la construction des sociomatrices. Le choix s'est porté sur l'utilisation uniquement des données connues (Little et Rubin 2002). Plusieurs résultats obtenus ne peuvent par conséquent être généralisés à l'ensemble de la population qu'avec un certain degré de précaution.

Il faut également rappeler que le nombre d'agents et de missions déployés par chaque souverain devait nécessairement être mis en parallèle avec leurs durées de règnes respectifs (Table 2). L'estimation de ces dernières a été cependant une tâche extrêmement malaisée. La plupart des dates retenues par l'historiographie se sont avérées inopérantes. Pour Louis II d'Italie, par exemple, on choisit habituellement pour sa durée de règne celle de sa possession du titre impérial, c'est-à-dire les années 855-875 (voir par exemple Zielinski 1991, col.2177 ; Contamine 2002, p.439). Sa première activité missatique pourtant a eu lieu en 845 (Manaresi 1955, n°49, pp.160-166 ; cité d'après Kikuchi 2013, pp.412-413 ; voir également Krause 1890, appendice II, n°139, p.294). L'intervalle durant lequel le fils de Lothaire faisait appel aux *missi* est ainsi plus espacé que sa durée de règne admise. Cette divergence s'explique en grande partie par la complexité de la situation politique en Europe occidentale durant la période étudiée. L'existence de plusieurs royaumes au début du IX$^e$ siècle, le morcellement de l'empire en entités politiques différentes à la deuxième moitié du même centenaire, la pratique des co-règnes entre le père et le fils ou bien encore des interruptions dans certains règnes (comme c'est cas de la courte abdication de Louis le Pieux) nous livre un tableau politique contrasté et difficile à saisir (Werner 1981 ; De Jong 2009 ; Goldberg 2006). La coexistence de plusieurs règnes explique également une discordance apparente entre la diminution, avec le temps, du nombre de missions (et d'agents) de chaque souverain et leur rythme soutenu, voire accéléré, tout au long du IX$^e$ siècle (Figure 1).

La question de la durée des règnes est simple et complexe à la foi : de quel règne s'agit-il ? Faut-il retenir pour Louis II d'Italie l'année 840, moment où il a entamé sa carrière royale en péninsule italienne ? Ou bien l'année 850 quand il a été couronné par le pape comme l'héritier de l'empire et a partagé le pouvoir avec son père Lothaire I encore sur le trône ? Ou plutôt, et c'est la date la plus souvent citée, son règne n'a-t-il commencé qu'en 855, moment à partir duquel il a été le seul à posséder le titre impérial ? Si son utilisation des *missi* a débuté en 845, c'est alors l'accession au trône du royaume de l'Italie (840) qui convenait au mieux pour cette étude. Les dates de début de chaque règne sont donc un sujet de débat et ont dû être examinées cas par cas. Leur choix a été dicté par l'accès, d'une façon ou d'une autre, au pouvoir royal et, par conséquent, par la possibilité de faire appel aux *missi dominici*. Il est donc impératif de souligner que ces dates ne sont proposées qu'à titre indicatif pour fournir un aperçu de la durée approximative durant laquelle chaque souverain pouvait employer des *missi*. (Pour toutes précisions sur les dates retenues voir les références dans Table 2, note 2).

En définitive, plusieurs types d'informations ont été recueillis dans le cadre de cette recherche. Pour les rois : nom et date du règne. Pour les *missi dominici* : fonction, lieu de fonction, mission, liens de parenté avec les autres envoyés ou avec les rois. Pour les missions : objectif, date et lieu. Enfin, pour chaque lieu mentionné on a collecté les coordonnées géographiques (latitude et longitude). Toutes les données ont été encodées et stockées dans une base de données relationnelle (MySQL). Le corpus contient ainsi des renseignements sur 14 rois/empereurs, 475 *missi dominici*, 398 missions et 307 lieux (Tables 1 et 2).

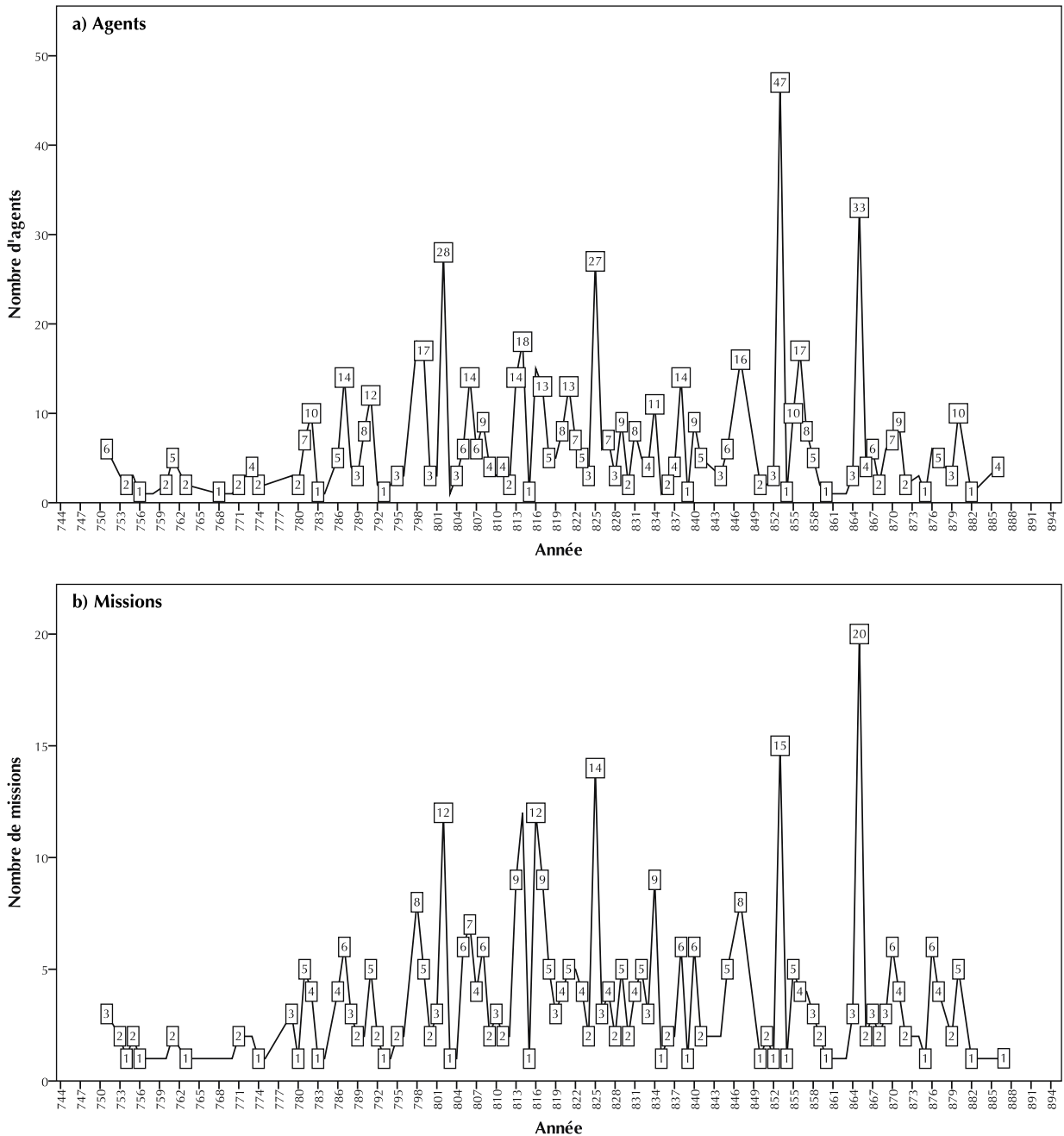

Figure 1 : Nombre des agents (*a*) et des missions (*b*) durant la période étudiée.

La présence des informations sur les fonctions des *missi* a permis par ailleurs d'ajouter des données supplémentaires dans la description des agents. Selon le rôle exercé (comte, évêque, archevêque, etc.) une catégorie (laïc ou ecclésiastique) a été attribuée à chaque *missus* recensé. Les premières observations laissent entrevoir une distribution quasi paritaire, aussi bien dans l'ensemble du réseau que parmi les agents employés par chaque roi, entre ces deux catégories (Table 3). Sur l'ensemble des 475 agents étudiés, 164 (34%) agents sont répertoriés comme « ecclésiastique », 147 (31%) comme « laïc », 4 (1%) ont une double catégorie et pour 160 (34%) agents cette information est inconnue.

|  | Durée de l'utilisation des *missi* (dates de première et de dernière mission)[1] | Durée approximative des règnes[2] | Nombre de missions[3] | Nombre d'agents (total)[4] | Nombre d'agents (uniques)[4] | Proportion d'agents uniques |
|---|---|---|---|---|---|---|
| **Pépin III le Bref** | 17 ans (751-768) | 751-768 | 15 | 25 | 21 | 0.84 |
| **Carloman I** | 1 an (763) | 754/768-771 | 1 | 2 | 2 | 1.00 |
| **Charlemagne** | 51 ans (763-814) | 754/768-814 | 129 | 231 | 153 | 0.66 |
| **Pépin d'Italie** | 3 ans (804-807) | 781-810 | 2 | 5 | 5 | 1.00 |
| **Louis le Pieux** | 45 ans (795-840) | 781/814-840 | 114 | 184 | 135 | 0.73 |
| **Pépin I d'Aquitaine** | 6 ans (828-834) | 814-838 | 2 | 2 | 2 | 1.00 |
| **Lothaire I** | 30 ans (823-853) | 814/843-855 | 20 | 38 | 33 | 0.86 |
| **Louis II de Germanie** | 43 ans (829-872) | 817/843-876 | 17 | 21 | 18 | 0.85 |
| **Charles II le Chauve** | 33 ans (844-877) | 843-877 | 57 | 118 | 83 | 0.7 |
| **Louis II d'Italie** | 29 ans (845-874) | 840/855-875 | 37 | 64 | 49 | 0.77 |
| **Lothaire II de Lotharingie** | 1 an (859) | 855-869 | 1 | 1 | 1 | 1.00 |
| **Carloman de Bavière** | 1 an (879) | 876-880 | 1 | 2 | 2 | 1.00 |
| **Louis II le Bègue** | 1 an (879) | 877-879 | 1 | 1 | 1 | 1.00 |
| **Charles III le Gros** | 10 ans (876-886) | 876/881-888 | 9 | 17 | 12 | 0.71 |

Table 2 : Résumé des données sur les missions, les agents et les souverains. (Dans l'ordre chronologique des débuts de règnes)

1 : Les dates correspondent à l'année de la première et de la dernière mission pour chaque roi.

2 : Les dates des règnes sont données à titre strictement indicatif. Si la date du premier accès au pouvoir correspond à une année autre que celle retenue habituellement, elle a été mise avant la barre oblique. Sur toutes les dates mentionnées voir en premier lieu les notices dans *Lexikon des Mittelalters*: (Zielinski 1991), (Störmer 1991a, 1991b), (Schneidmüller 1991a, 1991b, 1991c, 1993a, 1993b), (Jarnut 1991), (Goetz 1991a, 1991b), (Fleckenstein 1991a, 1991b, 1993) ainsi que (Vauchez 1997) et également les tables des règnes : (Contamine 2002, p.439) ; (Gauvard 2002, pp.527-528) ; (Riché 1997, pp.403-407).

Les précisions quant à la datation du premier accès au pouvoir royal :

Pépin III le Bref. Les dates correspondent à celles communément retenues. (Fleckenstein 1993 ; Gauvard 2002, p.527 ; Contamine 2002, p.439)

Carloman I, |754| - sacré par le pape Etienne II avec son frère Charlemagne. (Jarnut 1991, col.996)

Charlemagne, |754| - sacré par le pape Etienne II avec son frère Carloman I. (Fleckenstein 1991a, col.956)

Pépin d'Italie. Les dates correspondent à celles communément retenues. (Schneidmüller 1993a, col.2171 ; Contamine 2002, p.439)

Louis le Pieux, |781| - sacré par le pape Hadrien Ier en tant que roi d'Aquitaine. (Fleckenstein 1991b, col.2171)

Pépin I d'Aquitaine. Les dates correspondent à celles communément retenues. (Schneidmüller 1993b, col.2170 ; Contamine 2002, p.439)

Lothaire I, |814| - roi de Bavière. (Goetz 1991a, col.2123 ; cf. Contamine 2002, p.439 qui propose la date 817 correspondant à l'*Ordinatio Imperii* de Louis le Pieux et à association de Lothaire I au trône impérial)

Louis II de Germanie, |817| - roi de Bavière. (Störmer 1991a, col.2172 ; cf. Contamine 2002, p.439 qui propose la date de 814)

Charles II le Chauve. Les dates correspondent à celles communément retenues. (Schneidmüller 1991a, col.967 ; Gauvard 2002, p.528 ; cf. Contamine 2002, p.439 qui mentionne 840 comme date de début de règne)

Louis II d'Italie, |840| - roi d'Italie. (Zielinski 1991, col.2177)

Lothaire II de Lotharingie, |855| - roi de Lotharingie. (Goetz 1991b, col.2124)

Carloman de Bavière. Les dates correspondent à celles communément retenues. (Störmer 1991b, col.996 ; Contamine 2002, p.439)

Louis II le Bègue. Les dates correspondent à celles communément retenues. (Schneidmüller 1991b, col.2175-2176 ; Gauvard 2002, p.528 ; Contamine 2002, p.439)

Charles III le Gros, |876| - roi d'Alémanie. (Schneidmüller 1991c, col.968 ; cf. Contamine 2002, p.439 qui propose la date de 882 et Gauvard 2002, p.528 qui propose l'année 881 comme les dates de début de son règne)

3 : Le nombre des missions recensées est 398. Parmi elles 390 ont été commandées par 1 roi et 8 par 2 rois. Le nombre de toutes les missions de tous les rois présents dans cette colonne est obtenu comme suit : 390 (missions commandées par 1 roi) + 8x2 (missions commandées par 2 rois) = 406.

4 : Par *le nombre total des agents d'un roi*, on entend *tous les agents* qui ont été envoyés par ce roi ; le même agent pouvait être envoyé *plusieurs fois par le même roi*. L'abbé de Fontenelle Anségise, par exemple, a été sollicité en tant que *missus* de Louis le Pieux à trois reprises (Depreux 1997, notice 30, pp.104-106 ; Kikuchi 2013, pp.319-322) ; cet agent a été ainsi compté 3 fois dans la colonne « agents (total) » de Louis le Pieux. Par le *nombre des agents uniques* d'un roi, on n'entend que les agents *différents* qui ont été envoyés par ce roi. Anségise a été ainsi compté une seule fois dans la colonne « agents (uniques) » pour le règne de Louis le Pieux. Enfin, *par le nombre des agents répertoriés dans la base*, on entend le nombre des *agents différents qui ont existé durant la période étudiée tous rois confondus* (Table 1). Le même abbé de Fontenelle, hormis ses activités missatiques pour Louis le Pieux, a servi également une fois Charlemagne. Cet agent n'a été ainsi compté qu'une seule fois en tant qu'*agent répertorié dans la base*, bien qu'il soit présent aussi bien parmi les agents du premier empereur franc que parmi ceux de son fils.

|  | Agents uniques, (nb.)* | Ecclésiastiques, nb. (%) | Laïcs, nb. (%) | Mixte**, nb. (%) | Inconnu, nb. (%) |
|---|---|---|---|---|---|
| **Pépin III le Bref** | 21 | 9 (43%) | 4 (19%) | 0 | 8 (38%) |
| **Carloman I** | 2 | 2 (100%) | 0 | 0 | 0 |
| **Charlemagne** | 153 | 55 (36%) | 49 (32%) | 0 | 49 (32%) |
| **Pépin d'Italie** | 5 | 1 (20%) | 3 (60%) | 0 | 1 (20%) |
| **Louis le Pieux** | 135 | 47 (35%) | 47 (35%) | 1 (1%) | 40 (29%) |
| **Pépin I d'Aquitaine** | 2 | 0 | 0 | 0 | 2 (100%) |
| **Lothaire I** | 33 | 14 (43%) | 11 (33%) | 1 (3%) | 7 (21%) |
| **Louis II de Germanie** | 18 | 2 (11%) | 11 (61%) | 0 | 5 (28%) |
| **Charles II le Chauve** | 83 | 33 (40%) | 18 (22%) | 2 (2%) | 30 (36%) |
| **Louis II d'Italie** | 49 | 17 (35%) | 11 (22%) | 0 | 21 (43%) |
| **Lothaire II de Lotharingie** | 1 | 0 | 0 | 0 | 1 (100%) |
| **Carloman de Bavière** | 2 | 1 (50%) | 1 (50%) | 0 | 0 |
| **Louis II le Bègue** | 1 | 1 (100%) | 0 | 0 | 0 |
| **Charles III le Gros** | 12 | 4 (34%) | 7 (58%) | 0 | 1 (8%) |

Table 3. Distribution des *missi* laïcs et ecclésiastiques.
(Dans l'ordre chronologique des débuts de règnes, voir Table 2)

\* : Par les *agents uniques* on entend des agents *différents* employés par chaque roi (voir Table 2, note 4).
\*\* : Parmi les agents recensés, 4 ont endossé au cours de leur vie le statut de laïc et d'ecclésiastique. On peut citer par exemple Guy de Spolète qui après avoir été abbé devint duc, ou bien encore Hugues l'Abbé.

Quant aux objectifs des missions, bien que ces renseignements aient été retenus dans le corpus, leur analyse a été pour l'heure exclue de la présente étude. Un travail additionnel et approfondi les concernant fait partie des développements ultérieurs et il nécessitera sans doute le recours à des techniques autres que celles employées ici. On signalera uniquement que plusieurs pistes pour aborder les activités des *missi* dans le cadre de leur mission ont été déjà envisagées et ont donné des résultats prometteurs (Depreux 2001).

### 2.3 Modèles d'analyses

L'approche de l'analyse de réseau a exigé un regard attentif sur les différents types de données retenues. Si, dans une perspective métaphorique, il est possible de parler d'un seul et unique réseau aristocratique constitué des agents et des rois (Althoff 2004), dans le cadre méthodologique adopté il a été impératif de faire une distinction claire entre les différentes catégories des nœuds, les *modes*. Par un *mode*, on entend une collection des nœuds qui appartiennent au même type (Wasserman et Faust 1994, p.35 ; Newman 2010, pp.123-127). La nature des renseignements recueillis et les hypothèses de cette recherche ont dicté alors le choix des *modes* suivants : souverain, agent, mission. La complexité évoquée des données géographiques n'a pas permis de retenir également le « lieu » comme un *mode* supplémentaire. Dans le réseau construit, chaque *mode* a été relié par les missions commandées par le roi et effectuées par l'agent (Figure 2). Cependant l'analyse des relations entre ces trois *modes* a été difficile à appréhender dans le cadre d'un seul sociogramme. Le fait qu'un des *modes* est composé des missions (*événements*) a indiqué en outre qu'il s'agit d'un *réseau d'affiliation* (Wilson 1982 ;

Faust 1997) (Figure 3). La meilleure façon d'aborder ces interconnexions a été par conséquent d'établir plusieurs modèles d'analyse.

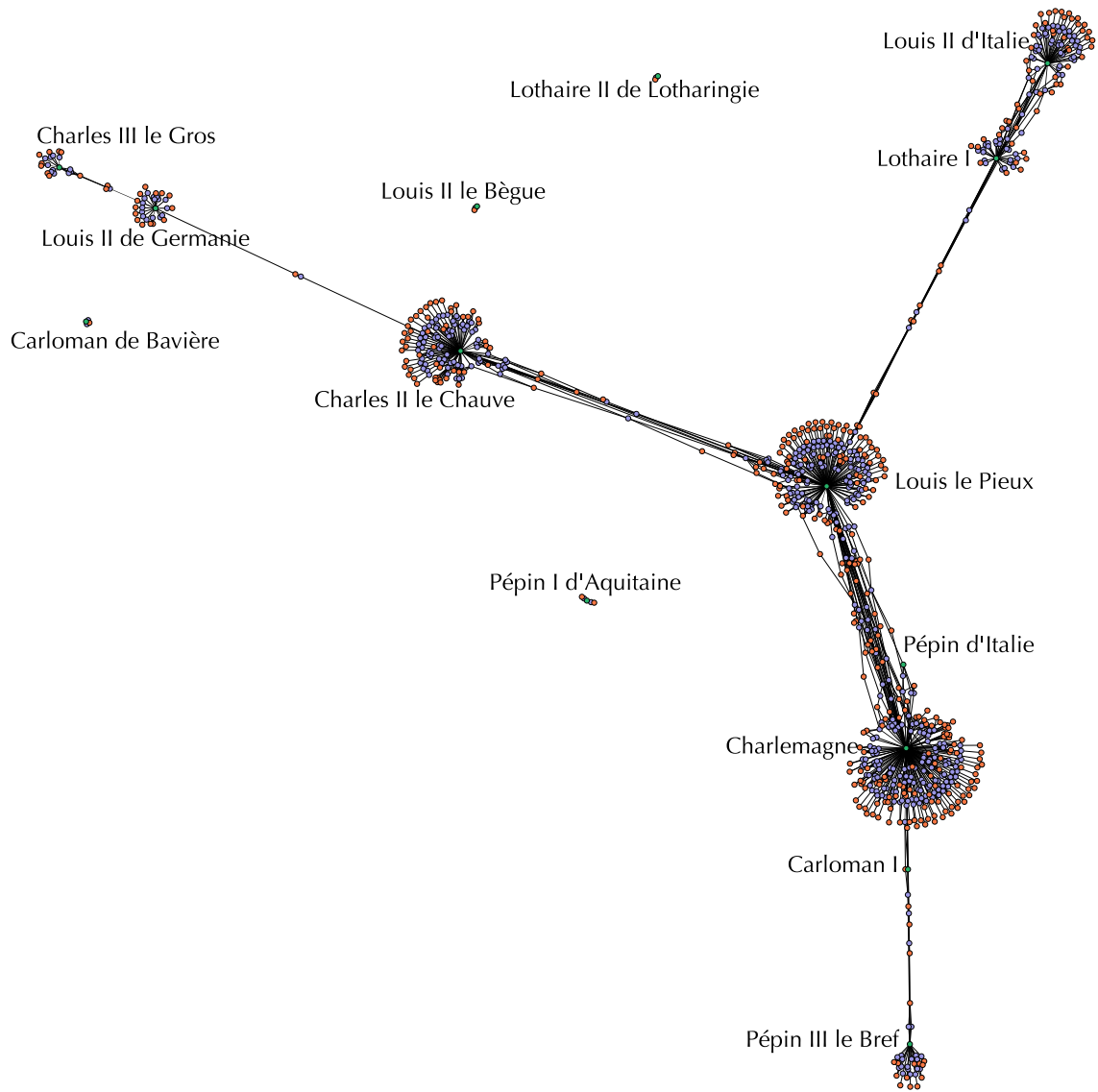

Figure 2 : Visualisation d'un réseau avec les nœuds (roi, agent, mission) reliés par les missions commandées et effectuées. (Couleurs des nœuds : vert - roi, bleu - agent, orange - mission).

*Réseau 1-mode (agent)*

La projection d'un événement (mission) sur l'acteur (agent) nous a permis d'obtenir un réseau 1-*mode* des agents reliés par les missions effectuées ensemble (lien non orienté et valué en fonction du nombre de missions en commun) (Table 4). Ce réseau a été construit pour des agents uniques de chaque roi (Table 2). Afin d'obtenir de meilleurs résultats, il était opportun de choisir les souverains ayant envoyé le plus grand nombre de *missi* : Pépin le Bref, Charlemagne, Louis le Pieux, Charles le Chauve, Lothaire I et Louis II d'Italie. La succession de ces six rois respecte par ailleurs l'évolution chronologique du système missatique durant les VIII$^e$-IX$^e$ siècles. Si la visée de ce modèle a été de mettre en évidence les groupes que les agents formaient au cours de leurs missions conjointes, leur comparaison pour les différents souverains a tenu à examiner les transformations qui y interviennent dans le temps. Afin d'observer plus en détail la composition de ces réseaux, le présent modèle a été enrichi par des données sur les catégories des agents (laïc, ecclésiastique).

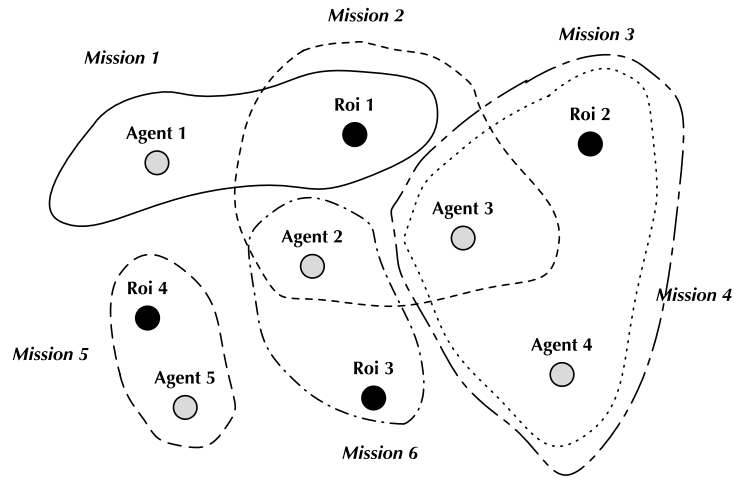

Figure 3 : Représentation graphique d'un réseau d'affiliation constitué de deux modes d'acteurs (roi, agent) et d'un mode d'événement (mission).

|   | **agent$_1$** | **agent$_2$** | **agent$_3$** | **agent$_4$** | **agent$_5$** |
|---|---|---|---|---|---|
| **agent$_1$** | - | 0 | 0 | 0 | 0 |
| **agent$_2$** | 0 | - | 1 | 0 | 0 |
| **agent$_3$** | 0 | 1 | - | 2 | 0 |
| **agent$_4$** | 0 | 0 | 2 | - | 0 |
| **agent$_5$** | 0 | 0 | 0 | 0 | - |

Table 4 : Sociomatrice d'un réseau 1-*mode* des agents, où le lien agent$_i$ → agent$_j$ est la mission effectuée ensemble.

*Analyse de positionnement multidimensionnel*

Outre les activités collégiales qui reliaient les *missi*, les données retenues ont permis d'examiner au plus près la proximité de leurs lieux de fonctions respectifs. À cette fin, il a été recouru au positionnement multidimensionnel classique à partir d'une matrice des distances euclidiennes (Table 5) (Cox et Cox 2001 ; Bartholomew *et al.* 2008). Chaque élément de matrice représente la distance la plus courte entre les lieux de fonctions des agents. Dans les cas où les *missi dominici* possédaient plusieurs lieux de fonctions, ce sont les endroits les plus proches l'un de l'autre qui ont été choisis pour le calcul. Le but de cette analyse a été d'observer les groupes que les agents pouvaient former selon l'éloignement entre leurs lieux de fonctions. Similaire au modèle de réseau 1-*mode*, le positionnement multidimensionnel a été réalisé pour six rois carolingiens ayant envoyé le plus grand nombre de *missi*. Les données sur les catégories des agents (laïc, ecclésiastique) ont été également introduites dans la visualisation.

|              | **agent$_1$** | **agent$_2$** | **agent$_3$** | **agent$_4$** | **agent$_5$** |
|---|---|---|---|---|---|
| **agent$_1$** | -    |      |      |      |   |
| **agent$_2$** | 121  | -    |      |      |   |
| **agent$_3$** | 129  | 427  | -    |      |   |
| **agent$_4$** | 138  | 1016 | 597  | -    |   |
| **agent$_5$** | 178  | 850  | 425  | 180  | - |

Table 5 : Matrice des distances euclidiennes (en km) utilisée pour l'analyse du positionnement multidimensionnel (les données utilisées sont celles des *missi* de Pépin le Bref, Figure 5a).

*Réseau 2-mode (agent, roi)*

Si les informations sur les relations entre les agents et les rois ont été pour une grande partie absentes (Table 1), un aperçu de la structure de l'institution missatique à travers les liens entre les *missi* et le pouvoir central a été indispensable. Le choix s'est porté sur l'examen d'un groupe spécifique des agents, ceux qui ont effectué des missions pour plusieurs souverains différents. L'objectif a été d'approcher le plus précisément possible le mécanisme de transition des agents d'un roi à l'autre et d'étudier son impact sur l'organisation générale du réseau. La sociomatrice ainsi construite comprend d'un côté des rois et de l'autre des *missi* déployés par plus d'un roi (lien orienté et valué en fonction du nombre de missions effectuées par agent pour chaque roi) (Table 6).

|              | **roi$_1$** | **roi$_2$** | **roi$_3$** | **roi$_4$** |
|---|---|---|---|---|
| **agent$_2$** | 1 | 0 | 1 | 0 |
| **agent$_3$** | 1 | 2 | 0 | 0 |

Table 6 : Sociomatrice d'un réseau orienté 2-*mode* (agents, roi), où le lien roi$_i$ → agent$_j$ est la mission commandée par un roi à un agent.

*Analyse descriptive*

L'examen des relations entre les agents et les lieux de leurs affectations a nécessité de faire appel à des techniques autres que celles d'analyse de réseaux. De fait, la spécificité de traitement des données géographiques déjà mentionnée plus haut rendait difficilement réalisable l'utilisation de ce type d'informations dans une sociomatrice ou un graphe. Ces relations ont été alors abordées par l'analyse des distributions des distances les plus courtes entre les lieux d'affectations d'un agent et ses lieux d'attaches personnelles ou de fonctions. Plusieurs unités d'observation ont été retenues. D'une part, pour examiner l'ensemble du réseau, on a pris d'abord la distance la plus courte entre les lieux d'affectations et les lieux de fonctions/d'attaches personnelles d'un agent. D'autre part, pour observer l'évolution chronologique de ce type de relations, les distances les plus courtes ont été regroupées par dates des affectations (Table 7). On note que d'une part, chaque agent pouvait avoir plusieurs lieux de fonction ou d'attaches personnelles et, d'autre part, chaque mission pouvait être commandée par plus d'un roi et avoir plus d'un lieu d'affectation. Par conséquent, dans l'étude, par exemple, des relations entre les lieux d'affectations et les lieux d'attaches personnelles d'un agent, on a calculé les distances pour chaque lieu d'affectation d'un agent et tous ses lieux de parenté. La distance la plus courte a été ensuite obtenue de l'ensemble

{distance1, distance2, …, distance$_k$}. Le même principe s'est appliqué pour l'étude des relations entre les lieux d'affectations d'un agent et ses lieux de fonctions.

| | Lieux de fonctions | Lieux d'attaches personnelles | Mission | | | Distance | |
| --- | --- | --- | --- | --- | --- | --- | --- |
| | | | Année | Roi-commanditaire | Lieux d'affectations | | |
| **agent** | lieu.fonction$_1$ | lieu.att.perso$_1$ | année | roi$_1$ | lieu.affectation$_1$ | distance$_1$ | distance la plus courte |
| | lieu.fonction$_2$ | lieu.att.perso$_2$ | | | lieu.affectation$_2$ | distance$_2$ | |
| | | | | roi$_2$ | | | |
| | lieu.fonction$_n$ | lieu.att.perso$_m$ | | | lieu.affectation$_l$ | distance$_k$ | |

Table 7 : Schéma des données utilisées pour l'analyse des relations entre l'agent et son lieu d'affectation.

Les visualisations et les indicateurs statistiques pour les analyses des réseaux (Figures 4, 5 et 6 ; Tables 8 et 9) ont été obtenus avec le logiciel *Gephi*. Pour la reproductibilité des résultats et la meilleure maîtrise des rendus visuels, les analyses de positionnement multidimensionnel ainsi que les analyses descriptives (Figures 7, 8, 9 et 10) ont été effectuées avec *R*.

## III RÉSULTAT

### 3.1 Caractéristiques et évolution du réseau des agents

Le premier volet des analyses porte sur l'examen de la structure du réseau missatique et de son évolution dans le temps à travers trois modèles : réseau 1-*mode* (agent), positionnement multidimensionnel et réseau 2-*mode* (agent-roi).

Dans le réseau 1-*mode* (Figure 4) la couleur des nœuds est définie par la catégorie à laquelle l'agent appartenait : rouge pour les ecclésiastiques, bleue pour les laïcs, verte pour les *missi* avec une catégorie double et grise si celle-ci est inconnue. Les liens qui relient les envoyés désignent les activités effectuées ensemble. Ces liens sont non orientés et valués en fonction du nombre des missions conjointes : plus ce nombre est élevé plus le lien est dense (Table 4). Si l'objectif de ce modèle est de mettre en lumière les missions communes comme un élément structurant du système missatique, la mise en perspective de réseaux de plusieurs souverains vise à observer si cette structure change dans le temps. Les six souverains retenus correspondent à ceux ayant le plus grand nombre de *missi* déployés (Table 2). Une grande différence dans le nombre de nœuds (agents) de chaque réseau appelle à être corrélée avec la durée de règne de chaque roi (Table 2). Il n'est guère utile, par exemple, de comparer l'ampleur du réseau de Charlemagne (Figure 4b) qui est resté au pouvoir plus de 45 ans avec celui de Pépin le Bref (Figure 4a) qui n'a possédé le titre royal que durant 17 ans. La fragilité, déjà évoquée plus haut, de l'estimation de l'étendue exacte de la détention du pouvoir rend difficile, voire impossible, une quelconque évaluation quantifiée de la relation entre le nombre de *missi* au service d'un souverain et la durée de règne de ce dernier.

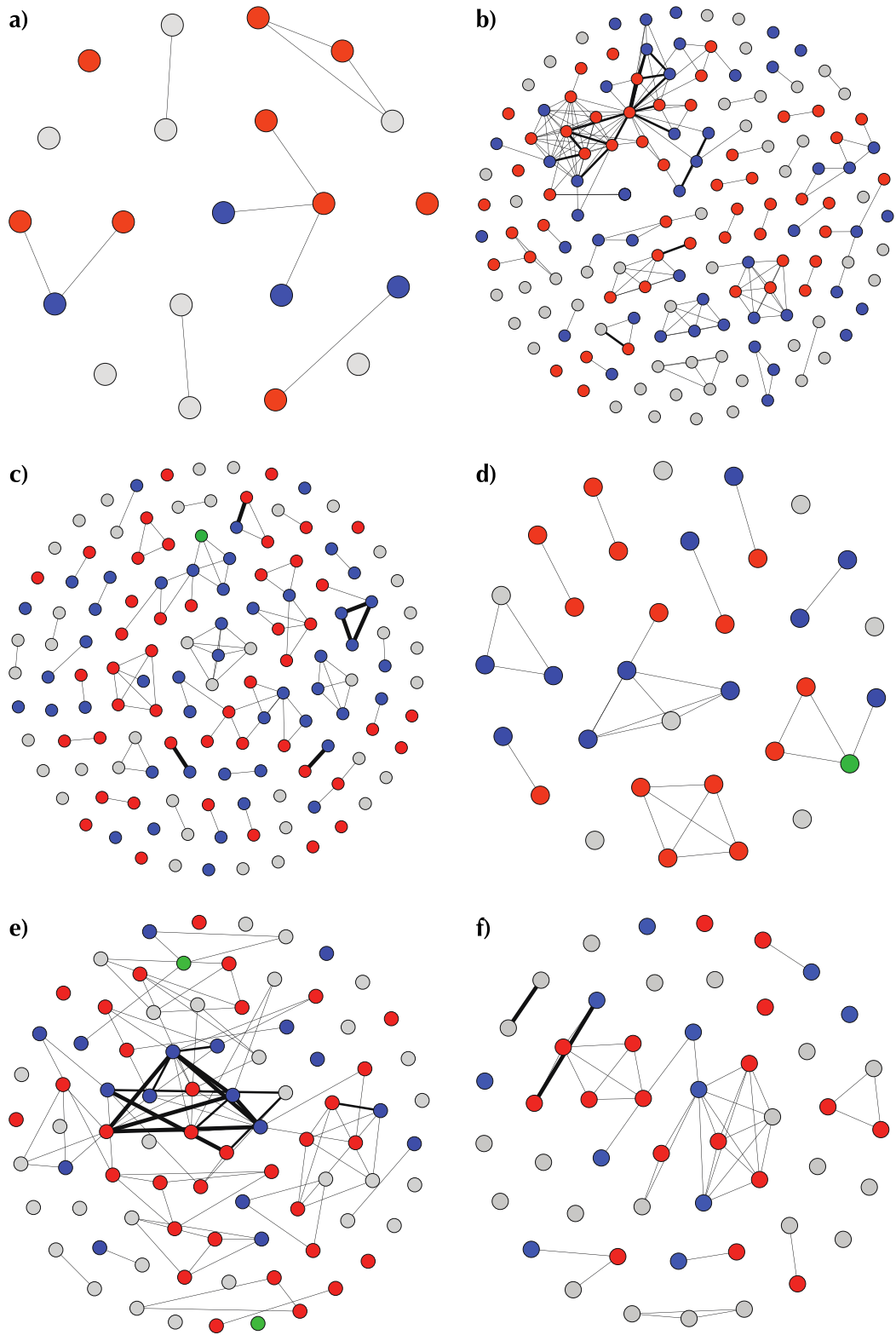

Figure 4 : Réseaux 1-*mode* des agents participant aux mêmes missions (couleurs : bleu - laïc, rouge - ecclésiastique, vert - mixte, gris - inconnu). (*a* - Pépin le Bréf ; *b* – Charlemagne ; *c* - Louis le Pieux ; *d*- Lothaire I ; *e* - Charles le Chauve ; *f* - Louis II d'Italie)

Les indicateurs statistiques livrent plusieurs éléments importants pour aborder l'examen de ces réseaux (Table 8). Une *composante connexe* désigne par exemple les nœuds interconnectés entre eux, mais non connectés aux autres composantes. Dans le cas des agents reliés par les activités missatiques, les composantes connexes sont les groupes formés par les agents qui ont exercé collégialement. Les nœuds isolés sont dès lors les *missi* qui n'ont fait aucune mission avec un autre agent. Un regard plus attentif sur la composition de ces groupes procure quelques indications sur la distribution des *missi* laïcs et ecclésiastiques. À l'exception de rares groupes, il ne semble pas toutefois exister un quelconque modèle apparent. La répartition générale des différentes catégories des agents aussi bien dans les composantes des réseaux que parmi les nœuds isolés paraît être aléatoire. Même si un pourcentage assez élevé des données inconnues sur les fonctions des agents nuit à la lecture complète de la structure des groupes, le résultat de ces observations nous renvoie en partie au constat déjà fait sur la distribution quasi paritaire des agents ecclésiastiques et laïcs dans les réseaux de tous les souverains (Table 3).

|  | Nb. de nœuds | Nb. de liens | Degré moyen | Densité du réseau | Nb. de composants connexes |
|---|---|---|---|---|---|
| **(a) Pépin le Bref** | 21 | 11 | 1.048 | 0.052 | 11 |
| **(b) Charlemagne** | 153 | 181 | 2.366 | 0.016 | 66 |
| **(c) Louis le Pieux** | 135 | 90 | 1.333 | 0.010 | 71 |
| **(d) Lothaire I** | 33 | 26 | 1.576 | 0.049 | 15 |
| **(e) Charles le Chauve** | 83 | 95 | 2.289 | 0.028 | 35 |
| **(f) Louis II d'Italie** | 49 | 42 | 1.714 | 0.036 | 24 |

Table 8. Indicateurs statistiques pour les réseaux 1-*mode* des agents.

Si l'étude des composantes est instructive à plusieurs égards, il sera cependant imprudent de comparer leur nombre entre les différents réseaux. En effet, ce chiffre doit être pondéré par le nombre de nœuds dans chaque graphe. Une fois de plus, il est peu étonnant que le réseau de Charlemagne (Figure 4b) possède plus de composantes que le réseau de Pépin le Bref (Figure 4a). Il est néanmoins possible de mettre ces informations en perspective grâce aux indicateurs de la densité (Table 8). La densité (Δ) se traduit alors comme un nombre de *liens présents* sur un nombre de *liens possibles* entre les nœuds. Elle est calculée comme suit (Wasserman et Faust 1994, formule 4.3, p.101) :

$$\Delta = \frac{2L}{g(g-1)} \qquad (1)$$

Où $L$ est le nombre de liens dans le réseau et $g$ est le nombre de nœuds. Plus cette valeur s'approche de 1, plus le réseau contient de liens et plus ses nœuds sont interconnectés. Pour les réseaux étudiés où les liens représentent les activités conjointes, la densité nous indique si les agents ont effectué les missions en commun avec le plus grand nombre des autres agents. Même si tous les réseaux ont une densité relativement faible, les réseaux de Pépin le Bref (Figure 4a), de Lothaire I (Figure 4d) et de Louis II (Figure 4f) se distinguent quelque peu par leur densité plus élevée et le nombre moindre de composantes. Il n'est pas imprudent d'en déduire que par rapport aux envoyés des autres réseaux la plupart des *missi* de ces trois rois ont effectué plus de missions ensemble. Là encore, il ne faut pas perdre de vue le nombre moins important des agents, en comparaison avec les autres graphes, contenus dans ces trois réseaux.

Une dernière observation des métriques porte sur le degré moyen de nœud. Cette mesure spécifie à combien d'autres nœuds, en moyenne, un nœud est connecté. Sa valeur ($\bar{d}$) est calculée comme suit (Wasserman et Faust 1994, formule 4.1, p.100) :

$$\bar{d} = \frac{2L}{g} \qquad (2)$$

Où $L$ est le nombre de liens dans le réseau et $g$ est le nombre de nœuds. Plus le degré de nœud est élevé, plus il possède de connexions. Dans les réseaux où les liens désignent les missions conjointes, la lecture de cette valeur montre avec combien d'autres agents en moyenne chaque *missus* a effectué de missions. Les degrés moyens pour les réseaux de Charlemagne (Figure 4b), de Charles le Chauve (Figure 4e), de Louis II (Figure 4f) et, dans une moindre mesure, de Lothaire I (Figure 4d) sont les plus élevés. Les *missi* de ces réseaux ont été le plus souvent déployés avec des agents *différents*. Ce constat semble se confirmer par l'observation visuelle ; les réseaux *b*, *e*, *f* et *d* livrent des structures relativement plus complexes.

En définitive, l'intégralité des métriques des réseaux présents mettent en avant ceux de Charles le Chauve (Figure 4e), de Lothaire I (Figure 4d) et de Louis II d'Italie (Figure 4f). Les réseaux missatiques de ces souverains sont les plus connectés et ils disposent du degré moyen et de la densité les plus élevés proportionnellement à leur nombre de nœuds. Ce constat nous amène à penser que les *missi* de ces réseaux ont effectué les missions avec le plus grand nombre d'autres agents.

Il serait toutefois hâtif de tirer des conclusions sur la structure des réseaux missatiques et leur évolution à partir uniquement des données sur les missions conjointes. D'autres types de liens et de relations reliaient les *missi* au service du pouvoir central. Althoff (2004) et plusieurs autres ont déjà montré que l'aristocratie altomédiévale formait un cercle relativement fermé fondé sur des interactions de dépendance, de domination ou d'obligation. Cependant, les sources ne nous laissent entrevoir qu'une partie infime de ces rapports complexes. L'ampleur extrêmement réduite des informations, par exemple, sur les liens de parenté des *missi* (seulement 14%, Table 1) rend illusoire toute tentative d'une analyse concluante de ce type de relation. Or, outre les délégations partagées, les *missi* recensés pour cette étude bénéficient d'un nombre important de renseignements sur leurs lieux de fonctions respectifs. Les étudier de plus près non seulement donne un aperçu de la spatialisation du système missatique, mais permet également de mettre en lumière d'autres types de relations qui pouvaient relier les agents. Le positionnement multidimensionnel déjà présenté plus haut (Figure 5), qui procède par le placement des points dans un espace à deux dimensions, a été à cet égard une des analyses les plus pertinentes. En fonction de la distance la plus courte entre leurs lieux de fonctions, les *missi* des réseaux précédents ont été positionnés sur les graphiques. Plus la distance entre les lieux des fonctions des agents a été courte, plus proches les agents se trouvent l'un de l'autre. Tout comme pour l'analyse des réseaux des missions conjointes, ce type de visualisation vise à mettre en lumière les groupes que les agents pouvaient former selon la proximité de leurs lieux de fonctions. Cependant, en raison des informations manquantes, cette analyse ne pouvait être effectuée que pour les agents ayant un lieu de fonction connu. Une partie des agents des réseaux précédemment analysés (Figure 4) ont été par conséquent exclus. Bien que ce type de visualisation garde les distances réelles, exprimées en kilomètres, et tente de respecter la disposition géographique des lieux (les lieux situés plus au nord sont en haut de chaque visualisation, les lieux situés plus au sud sont en bas, etc.), ces représentations ne sont qu'une approximation de la carte géographique réelle et les points zéro varient d'un graphique à l'autre.

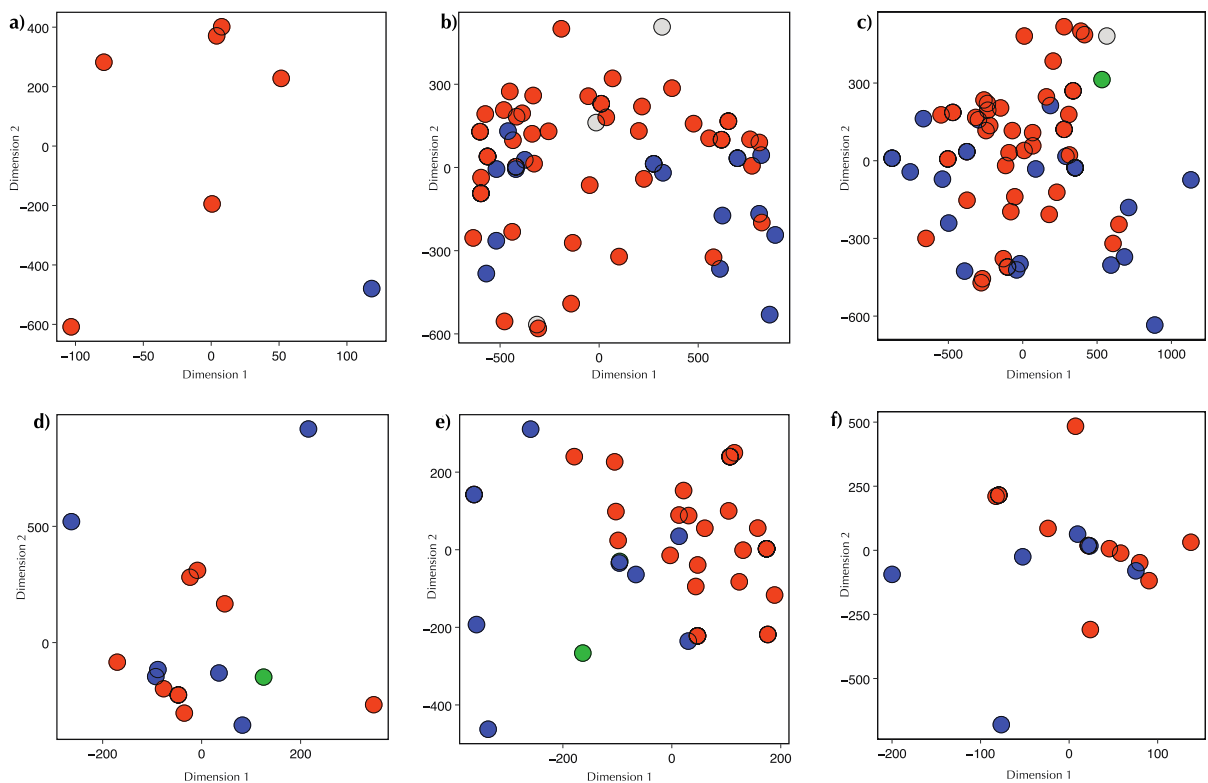

Figure 5 : Positionnement multidimensionnel des agents selon la distance la plus courte (en km) entre leurs lieux de fonctions (couleurs : bleu - laïc, rouge - ecclésiastique, vert - mixte, gris - inconnu). (*a* - Pépin le Bréf ; *b* – Charlemagne ; *c* - Louis le Pieux ; *d* - Lothaire I ; *e* - Charles le Chauve ; *f* - Louis II d'Italie)

En définitive, cette analyse permet d'examiner plusieurs nouvelles caractéristiques des réseaux étudiés. Le premier survol des graphiques est néanmoins peu concluant : aucune structure apparente ne se dégage. De même que pour les réseaux précédents les *missi* disposés selon leurs lieux de fonctions ne semblent pas former des groupes spécifiques et la répartition des agents suivant leurs catégories (laïc, ecclésiastique, mixte, inconnu) est également incertaine (cf. Bougard 1995, p.178). Or, la comparaison de l'ampleur géographique des différents réseaux livre plusieurs informations intéressantes sur le cadre spatial des lieux de fonctions des *missi* engagés. Mis à part les envoyés de Pépin le Bref (Figure 5a), ce sont les agents de Charlemagne (Figure 5b) et de Louis le Pieux (Figure 5c) qui ont été recrutés dans un rayon assez important de plus de 1000 kilomètres (cf. McKitterick 2008, pp.178-186, notamment p.184). Les cas de Nibridius (Kikuchi 2013, pp.546-549) et de Hildebaud (Depreux 1997, notice 151, pp.246-247 ; Kikuchi 2013, pp.458-460) qui ont servi tous les deux aussi bien le premier empereur franc que son fils donnent une perception de l'étendue de cette distance. Si le premier s'acquittait de ses fonctions d'archevêque à Narbonne, Hildebald, quant à lui, exerçait à Cologne ; les deux villes étant séparées par plus de 1100 kilomètres. Le territoire à partir duquel les *missi* ont été appelés est cependant réduit de presque deux fois durant les règnes de Lothaire I, de Charles le Chauve et de Louis II d'Italie (Figures 5d, 5e et 5f). Deux *missi*, Adalgis, comte de Spolète (Kikuchi 2013, pp.270-273), et l'évêque de Modène, Walpert (Kikuchi 2013, pp.644-645) auxquels recourut Louis II d'Italie témoignent bien de cette diminution de la distance entre les lieux de fonctions des agents mobilisés (Adalgis, point bleu, et Walpert, point rouge, sont situés à l'extrémité de l'axe X du Figure 5f). Les deux endroits, Spolète et Modène, logés sur les confins du cadre de recrutement de Louis II, ne sont séparés que par environ 300 kilomètres.

La comparaison de l'agencement des zones à partir desquels des *missi* ont été enrôlés par Charlemagne (Figure 5b) et Louis le Pieux (Figure 5c) révèle en outre des schémas différents. Si pour le premier on note une dispersion relativement paritaire des agents avec l'embryon de deux groupes possibles (à gauche et à droite), le deuxième nous montre un ensemble moins polarisé. Les *missi* sous l'autorité de Charlemagne ont été, semble-t-il, recrutés davantage dans les aires périphériques tandis que ceux de son fils ont plutôt exercé leurs fonctions au cœur des territoires impériaux (cf. Werner 1980, pp.205, 210).

Il est néanmoins nécessaire de garder une certaine modération à l'égard des résultats fournis par le positionnement multidimensionnel. L'évolution constante de la topographie du pouvoir, affectée par des partages territoriaux et des successions, est difficile à appréhender dans le cadre de cette analyse. Tantôt scindés, tantôt assemblés, les royaumes et leurs frontières n'ont été guère statiques dans la deuxième moitié du IX$^e$ siècle.

Tout compte fait, il est imprudent de se limiter, dans l'examen du réseau missatique, à la seule étude des agents. Les relations qu'ils entretenaient avec le pouvoir central ont joué un rôle important dans la construction de l'institution. Cependant, les données disponibles restreignent considérablement la palette des relations possibles à examiner. Toute tentative d'appréhender le rapport entre les rois et leurs envoyés à travers, par exemple, des liens de parenté est compromise : seulement 14% de cette information est connue (Table 1). Il est toutefois possible de cerner quelques nouveaux aspects de leurs interconnexions par le biais des missions qui reliaient les souverains et leurs *missi*. L'observation d'un réseau composé des agents, des rois et des missions (Figure 2) nous a déjà livré un tableau chronologique assez fidèle de l'ensemble du système missatique où chaque souverain représentait une étape importante dans l'évolution du réseau. Cette visualisation a laissé également entrevoir le mécanisme de transition des agents d'un règne à l'autre. Un modèle plus précis d'un réseau 2-*mode* (agent, roi) déjà présenté plus haut (Table 6) invite à examiner ce processus de plus près (Figure 6). Ce réseau est constitué des *missi* ayant servi plus d'un roi et des souverains (48 nœuds, dont 10 rois et 38 agents ; les rois suivants ne disposaient d'aucun *missus* ayant servi plus d'un souverain : Pépin I d'Aquitaine, Lothaire II, Carloman de Bavière, Louis II le Bègue). Si l'orientation des liens désigne les missions commandées par un souverain à un *missus*, leur épaisseur indique le nombre de ces missions.

Le premier constat est immédiat : Charlemagne et Louis le Pieux ont eu le plus grand nombre de *missi* en commun. Rappelons-le toutefois : Charlemagne et son fils disposaient de la quantité la plus importante de missions commandées et d'agents déployés. Leurs durées de règnes respectifs sont également parmi les plus longues (Table 2). L'observation suivante confirme les conclusions antérieures (Figure 2) : deux branches se distinguent clairement, d'un côté Charles le Chauve, Louis II de Germanie et Charles III le Gros, de l'autre, Lothaire I et son fils Louis II d'Italie. Cette disposition, tout comme le dessin général du réseau, semble se conformer à la succession des règnes. De fait, après la mort de Pépin le Bref en 768, ce sont ses fils Charlemagne et Carloman I qui ont codirigé brièvement le royaume jusqu'à la mort du dernier en 771. Pépin d'Italie, le fils de Charlemagne, est monté sur le trône des territoires apennins en 781 encore très jeune et y est resté jusqu'à sa mort en 810. Quant à Louis le Pieux, un autre fils de Charlemagne, c'est lui qui a hérité en 814 de l'empire de son père. La succession de Louis a connu cependant plusieurs tourmentes et ce n'est qu'en 843 lors du partage de Verdun que ses fils sont arrivés à cosigner un accord. Deux camps s'y sont retrouvés : Charles le Chauve et Louis II de Germanie d'un côté et Lothaire I de l'autre. Au fil des legs et des partages, le fils aîné de ce dernier, Louis II d'Italie a hérité de son père d'abord le royaume d'Italie et ensuite le titre impérial. Le fils de Louis II de Germanie, Charles III le Gros est parvenu toutefois, vers la fin de sa vie, à cumuler plusieurs titres royaux et à restaurer peu ou prou l'espace impérial unifié du début du IX$^e$ siècle.

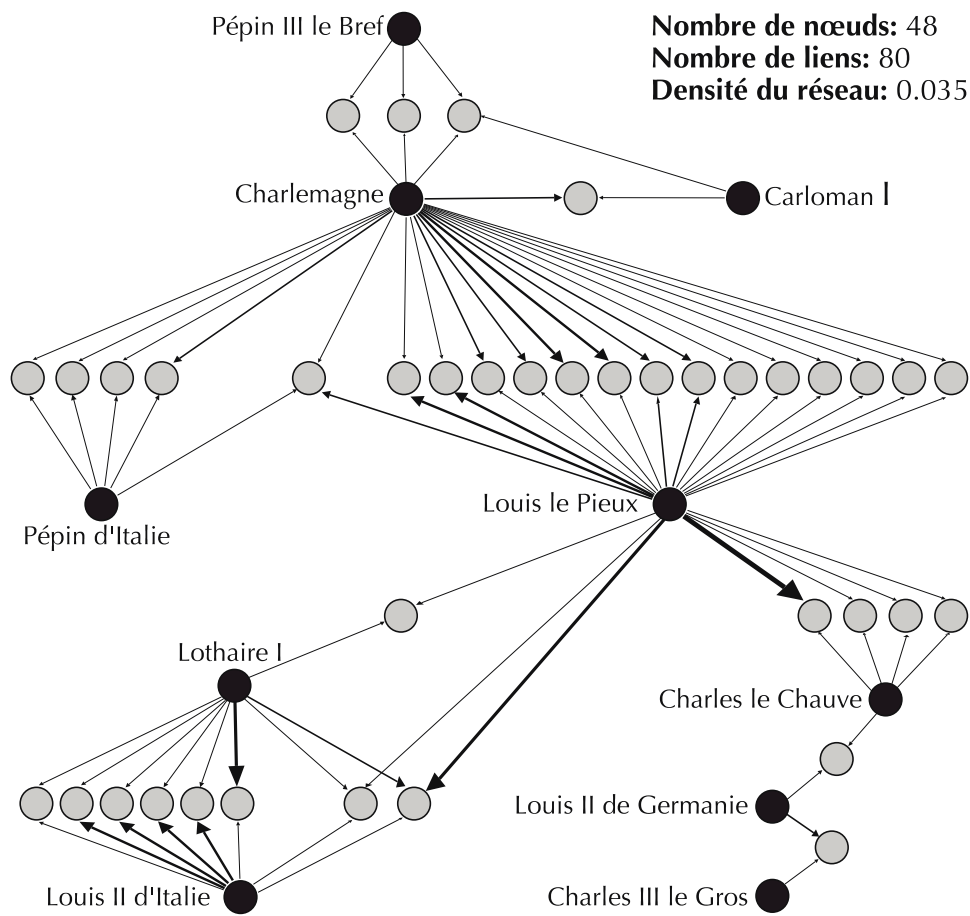

Figure 6 : Réseau orienté 2-*mode* (agent, roi) avec des agents ayant servi plus d'un roi.

|  | **Degré sortant** | **Degré sortant pondéré** | **Centralité intermédiaire** |
|---|---:|---:|---:|
| **Pépin III le Bref** | 3 | 3.0 | 0.0020 |
| **Carloman I** | 2 | 2.0 | 0.0006 |
| **Charlemagne** | 23 | 33.0 | 0.4200 |
| **Pépin d'Italie** | 5 | 5.0 | 0.0090 |
| **Louis le Pieux** | 22 | 37.0 | 0.6400 |
| **Lothaire I** | 9 | 13.0 | 0.1400 |
| **Louis II de Germanie** | 2 | 3.0 | 0.0800 |
| **Charles II le Chauve** | 5 | 5.0 | 0.1600 |
| **Louis II d'Italie** | 8 | 16.0 | 0.0900 |
| **Charles III le Gros** | 1 | 1.0 | 0.0000 |

Table 9. Indicateurs statistiques pour le réseau 2-*mode* (agent, roi). (Dans l'ordre chronologique des débuts de règnes, voir Table 2).

Pour les métriques de ce réseau trois indicateurs ont été retenus : degré sortant, degré sortant pondéré et centralité intermédiaire.

Le premier, degré sortant, désigne, pour les réseaux orientés, le nombre de liens partant du nœud. Cela indique, dans notre cas, à combien de *missi* chaque roi a fait appel. Charlemagne et son fils, avec leurs degrés respectifs de 23 et 22, possèdent une partie dominante de tous les agents présents. Mis en perspective avec le nombre total de leurs *missi dominici* ce constat est toutefois peu surprenant (Table 2).

Le degré sortant pondéré, quant à lui, propose une mesure qui « pondère » le nombre des liens partant du nœud par la valeur desdits liens. Comme déjà évoqué, la valeur des liens correspond pour ce réseau au nombre des missions confiées à l'agent. Le degré sortant pondéré permet ainsi de prendre en compte l'intensité des liens entre le souverain et ses *missi*. Par cette mesure Louis le Pieux dépasse non pas uniquement son père (37 contre 33) mais, et de loin, tous les autres rois. Le fils de Charlemagne a donc été le plus apte à commander des missions aux agents ayant exercé pour d'autres rois. C'est, par exemple, le cas pour Donat, comte de Melun, qui a effectué 7 missions, dont 6 pour Louis le Pieux et une pour Charles le Chauve (Depreux 1997, notice 74, pp.160-162 ; Kikuchi 2013, pp.368-371). On peut de même mentionner l'abbé Anségise de Fontenelle ayant servi comme *missus* une fois Charlemagne et à trois reprises Louis le Pieux (Depreux 1997, notice 30, pp.104-106 ; Kikuchi 2013, pp. 319-322).

Le dernier indicateur, centralité intermédiaire, cherche à évaluer l'importance (*centralité*) d'un nœud dans le réseau.[3] Plus souvent un nœud se trouve « intermédiaire » sur le « chemin » entre deux autres nœuds, plus cette mesure est élevée. Dans le réseau étudié c'est Louis le Pieux qui possède la centralité intermédiaire la plus élevée (0.64) non pas uniquement vis-à-vis de son père Charlemagne (0.42) mais aussi à l'égard des autres rois. Parmi tous les souverains, Charles le Chauve et Lothaire I n'arrivent pas à franchir le seuil de deux dixièmes (0.14 pour Lothaire I et 0.16 pour Charles le Chauve). L'héritier du premier empereur franc confirme ainsi son rôle en tant que point d'articulation important dans le réseau de transition des agents entre les différents règnes. Il suffit en effet d'enlever Louis le Pieux du graphe pour constater que ce dernier se fractionnera en trois groupes distincts : Pépin le Bref, Carloman I, Pépin d'Italie et Charlemagne d'un côté, Lothaire I et Louis II de l'autre, et enfin Charles le Chauve, Louis II de Germanie et Charles le Gros.

Une certaine prudence est néanmoins de mise face aux conclusions qui ont pu être tirées à partir de l'analyse de ce réseau. Le nombre d'agents ayant servi plusieurs souverains est peu élevé : sur l'ensemble de 475 *missi*, seulement 38 ont été envoyés par plus d'un souverain (dont 34 *missi* ayant servi 2 souverains différents et 4 *missi* ayant servi 3 souverains différents).

### 3.2 Relations entre les agents et leurs lieux d'affectations

La deuxième partie de cette étude vise à répondre aux questions du recrutement des *missi* au sein de l'aristocratie locale. Les analyses des relations entre les lieux d'affectations des agents et leurs lieux de fonctions ou d'ancrages familiaux cherchent à évaluer la proximité de ces différents endroits entre eux. L'estimation des distances géographiques, bien qu'elle puisse paraître quelque peu réductrice au point de vue historique, apporte néanmoins plusieurs renseignements importants. Le recrutement des *missi* dans les zones proches de leur affectation a un impact manifeste sur la logique de gouvernance des territoires par le pouvoir central (Hannig 1984b). Il est toujours utile de souligner que les résultats obtenus sont influencés par une proportion importante d'informations manquantes. Les données recueillies ne fournissent que 44% de renseignements sur les lieux de fonctions des agents et 14% sur leurs attaches personnelles (Table 1).

---

[3] Bien que la centralité intermédiaire puisse être calculée de façons différentes pour les graphes orientés (White et Borgatti 1994 ; Faust 1997), dans le cas présent elle est évaluée pour les liens pris comme non orientés.

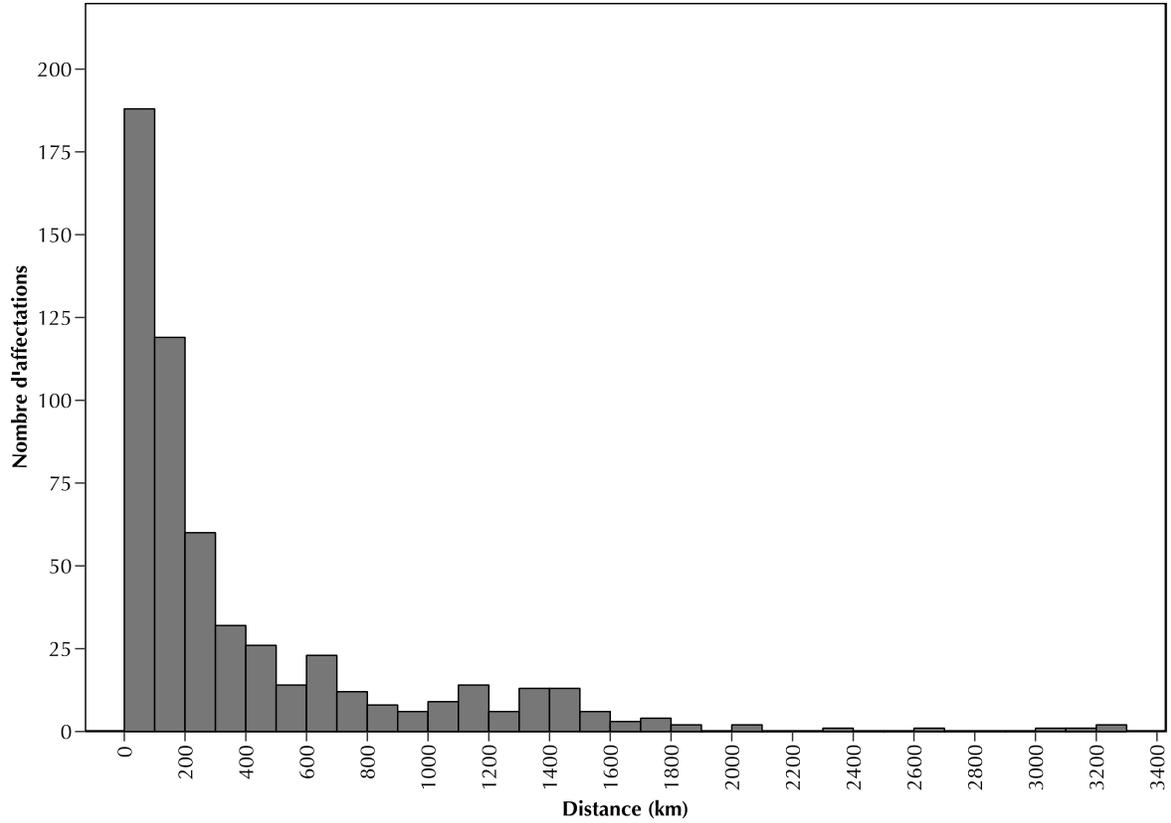

Figure 7 : Distribution des affectations par la distance la plus courte entre le lieu de fonction d'un agent et son lieu d'affectation. (Le nombre d'affectations analysées est de 566)

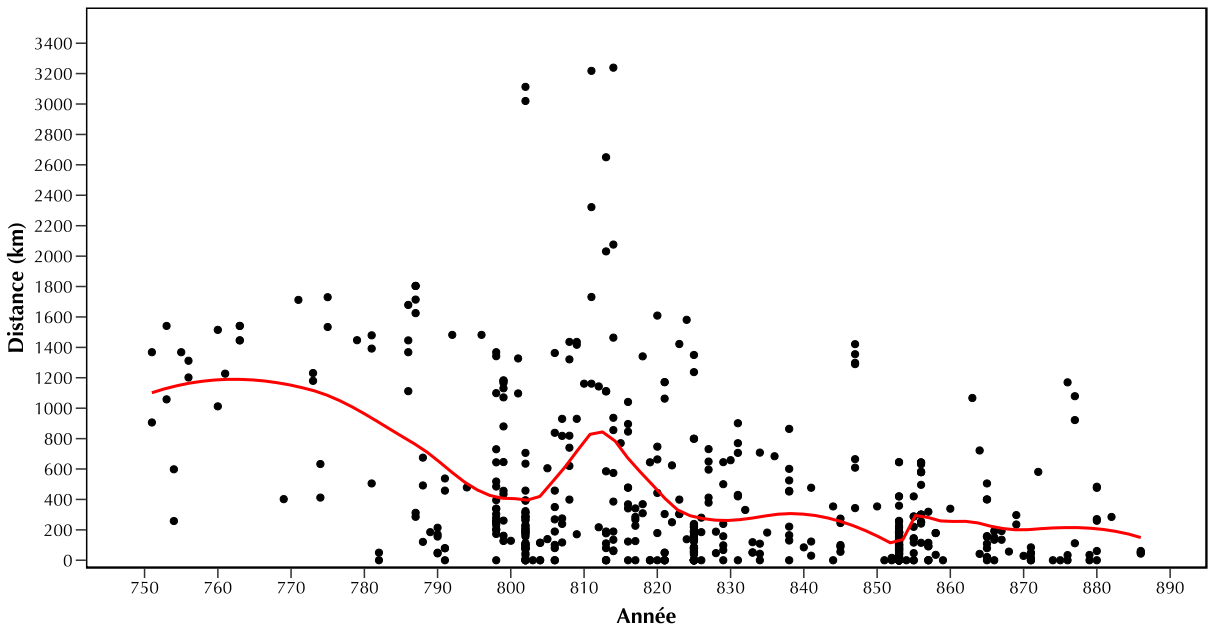

Figure 8 : Évolution de la distance la plus courte entre le lieu de fonction d'un agent et son lieu d'affectation. (Le nombre d'affectations analysées est de 566)

Le premier graphique livre la distribution des affectations par la distance la plus courte entre le lieu d'affectation et le lieu de fonction des *missi* (Figure 6). Dans les cas, fréquents, où plusieurs lieux de fonctions sont connus pour l'agent, cette distance a été calculée pour le lieu de fonction le plus proche de l'endroit de l'affectation (Table 7).

De façon chiffrée, l'analyse statistique confirme ainsi l'hypothèse du recrutement des agents sur les territoires proches de ceux où ils ont été envoyés pour des activités missatiques. Un tiers des affectations (environ 200 sur un nombre total de 566 affectations analysées) ont eu lieu dans un rayon de 100 kilomètres à partir des endroits où l'agent était déjà en exercice. En même temps, dans quelques rares cas, on constate des distances considérables (plus de 3000 kilomètres) séparant les endroits des missions des agents de leurs lieux de pouvoir. Le problème, déjà évoqué, de la dénomination des différents types d'envoyés et de l'emploi du vocable *missi* pour les délégations diplomatiques à l'étranger explique aisément un tel éloignement. L'exemple de Hugo qui a servi Charlemagne le démontre bien. Ce *missus* ayant des charges comtales à Tours s'est vu confier en 811 une mission, avec deux autres agents, à Constantinople (*Annales regni Francorum*, a. 811, p.133 ; Kikuchi 2013, p.295). Une autre illustration de distances parfois longues parcourues par les *missi* est fournie par la délégation commandée en 814/815 par Louis le Pieux. Ricouin, comte de Poitiers, a été envoyé, comme Hugo quelques années avant lui, du coeur des terres franques à la capitale de l'empire byzantin (*Annales regni Francorum*, a. 814, pp.140-141 ; Depreux 1997, notice 233, p.365 ; Kikuchi 2013, p.551 ; sur le lieu de fonction de ce comte, voir Hlawitschka 1960, p.296).

La mise en perspective temporelle des distances entre le lieu de fonction d'un agent et son lieu de mission offre la possibilité d'étudier au plus près l'évolution chronologique de cette relation (Figure 7). Appuyé par l'analyse statistique, le graphique apporte plusieurs renseignements intéressants. Tout d'abord, la corrélation entre les deux variables (la distance et l'année d'affectation) est de -0.428 avec p-valeur égal à 0. Il n'est pas imprudent d'en conclure alors que la diminution de la distance est corrélée à l'augmentation de l'année. Le trajet parcouru par les *missi* entre leur assise locale et l'endroit de leur affectation s'est réduit au fil du IX$^e$ siècle. Or, rien ne nous indique de façon précise quelle a été la cause exacte de ce phénomène. En dépit d'un résultat statistiquement significatif (p-valeur moins de 0.0005), la régression linéaire pour les relations entre les variables « année » et « distance » reste peu élevée ($R^2$=0.183). Le temps n'est qu'un des paramètres pour expliquer la baisse de la portée géographique des activités missatiques.

Le lissage (ligne rouge) présent sur le graphique apporte la possibilité d'un suivi plus détaillé de l'évolution chronologique des distances entre les lieux d'affectations et les lieux de fonctions des agents. Utilisée souvent en analyse des données longitudinales, cette mesure dresse une courbe en fonction de la moyenne calculée sur les valeurs avoisinantes (voir par exemple Simonoff 1996). Son tracé, bien qu'il semble confirmer la tendance de régression déjà constatée, nuance néanmoins les faits ; le mouvement n'est guère rectiligne et un bond important se dessine pour les années 800-815. Ce bond correspond aux missions à Constantinople déjà mentionnées pour les années 802, 811, 813 et 814 (dont trois envoyées par Charlemagne et une par Louis le Pieux) (Depreux 1997, pp.234, 243, 262, 408 ; Kikuchi 2013, pp.295, 307-308, 451, 551).

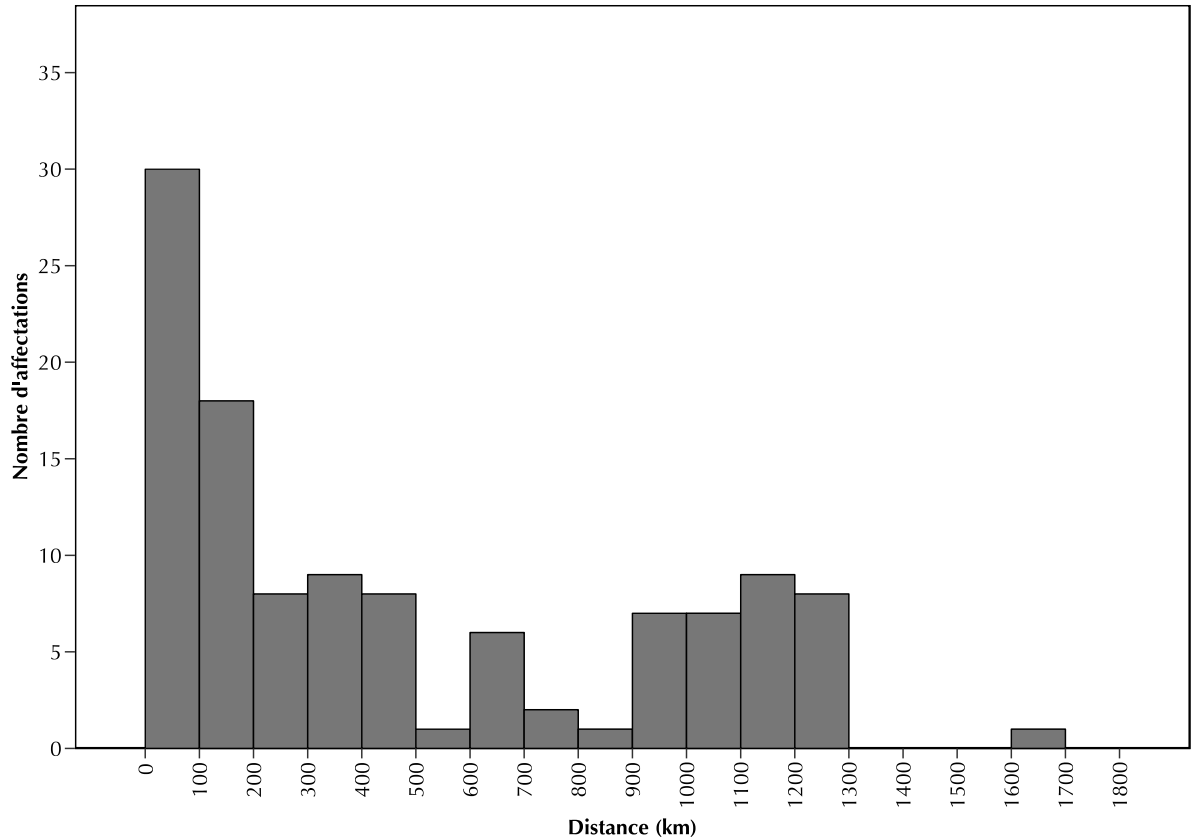

Figure 9 : Distribution des affectations par la distance la plus courte entre le lieu d'attaches personnelles d'un agent et son lieu d'affectation. (Le nombre d'affectations analysées est de 155)

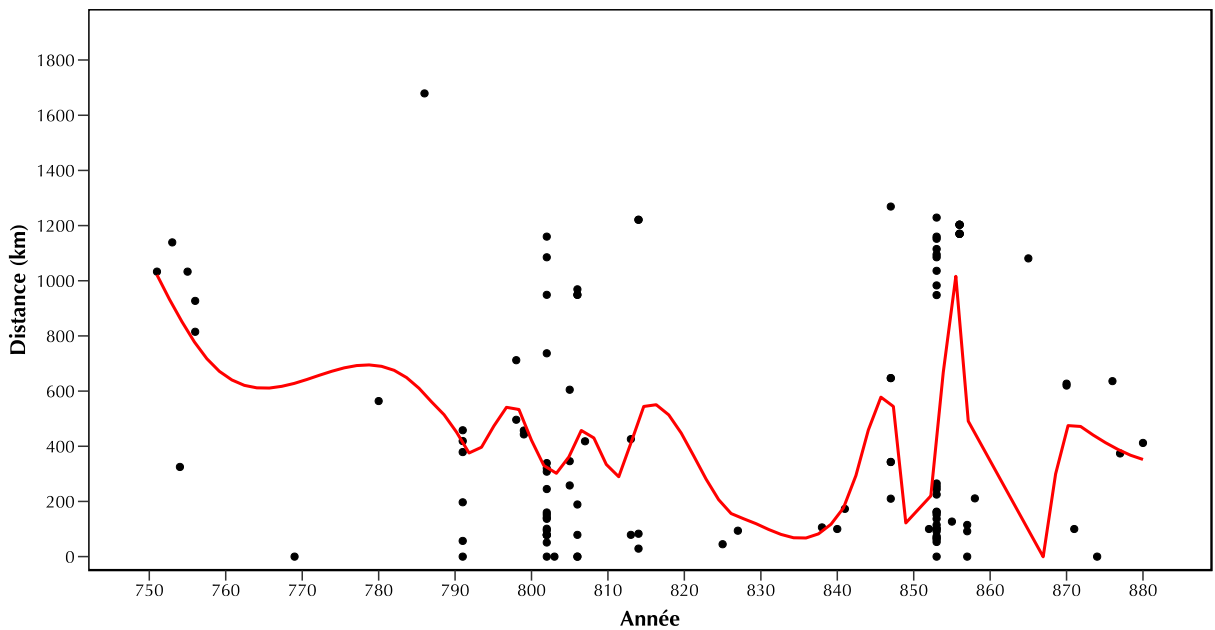

Figure 10 : Évolution de la distance la plus courte entre le lieu d'attaches personnelles d'un agent et son lieu d'affectation. (Le nombre d'affectations analysées est de 155)

Bien que ces analyses soient en grande partie affectées par un nombre important de données manquantes, il est également opportun d'avoir un aperçu de la distance entre les lieux d'affectations des *missi* et leurs lieux d'attaches personnelles. Le premier constat semble corroborer les conclusions antérieures : une partie non négligeable des affectations a eu lieu dans un rayon de 100 kilomètres autour des endroits d'attaches personnelles des *missi* (30 affectations sur 155 analysées) (Figure 8). La lecture du graphique du changement dans le temps de cette distance est marquée sans surprise par le caractère lacunaire des renseignements. De même, les analyses statistiques sont peu éloquentes : la régression linéaire et la corrélation (variables « distance » et « année ») ont une p-valeur supérieure à 0.05. Le tracé confus de lissage amène à une conclusion similaire.

En dépit de ces faiblesses apparentes, les analyses de la distribution de la distance la plus courte entre les lieux d'affectations d'un agent et ses lieux d'attaches personnelles fournissent quelques indications dignes d'intérêt. Les trajets les plus longs effectués par les *missi* depuis leurs lieux d'origines géographiques ou endroits où ils avaient des liens de parenté ne dépassaient que rarement le seuil de 1200-1300 kilomètres. Cet éloignement s'est tout particulièrement accentué vers la fin du VIII$^e$ siècle, le début et le milieu du siècle suivant. Quelques activités missatiques en témoignent fort bien. Le comte Audulf, originaire, en toute vraisemblance, de la noblesse bavaroise, a été envoyé par Charlemagne en 786 pour une mission militaire en Bretagne (*Annales regni Francorum*, a. 786, p.72 ; Krause 1890, appendice II, n°19, p.283 ; Kikuchi 2013, p.335). Erlwin, le comte palatin issu de la famille Widonides native d'Austrasie et implantée par la suite en Italie, intervient en tant que *missus* de Louis II de Germanie en Rhin moyen au cours de l'année 865 (DD LD, n°117, pp.166-167 ; Krause 1890, appendice II, n°199, p.299 ; Kikuchi 2013, p.384). L'exemple du comte de Paris Gérard est également évocateur. Apparenté, par son mariage avec Bertha, la sœur d'Ermengarde de Tours, à la famille Étichonides, avec leur forte assise alsacienne, Gérard s'est retrouvé, lors de son unique mission en 847, en Italie méridionale (Conc. 3, n°12, pp.133-139, ici page 139 ; Krause 1890, appendice II, n°138, p.294 ; Kikuchi 2013, pp.420-421). Quoi qu'il en soit, le faible nombre de données connues et le caractère peu significatif des tests statistiques nous invitent à garder une grande prudence devant les résultats de ces deux dernières analyses.

## IV   DISCUSSION ET CONCLUSION

Cette recherche a essayé de porter un regard nouveau sur les agents du pouvoir central en Europe occidentale du premier Moyen Âge. Les premières observations se sont concentrées sur la structure du réseau missatique et les changements qui s'y sont opérés durant plus d'un siècle. Les questions de rotation des agents et des groupes qu'ils pouvaient former au fil des missions conjointes semblent quelque peu avoir échappé à la loupe des enquêtes historiques précédentes (cf. toutefois Eckhardt 1956, p.516 ; Werner 1980, p.203). Ces interrogations sont pourtant inévitables pour saisir les traits organisationnels d'une institution. Un regard attentif sur ce mécanisme des activités collégiales a permis de toucher au plus près, chiffres à l'appui, à cet aspect de l'architecture des réseaux missatiques de six souverains carolingiens (Figure 4 ; Table 8). Tout compte fait, ce sont les réseaux de Charles le Chauve, Lothaire I et Louis II d'Italie qui ont été les plus structurés. La rotation des envoyés spéciaux a été moindre et la plupart d'entre eux ont effectué des missions ensemble. Cela laisse à penser que le pouvoir central a commencé à s'appuyer sur les mêmes agents, ce qui, si ce n'est pas un signe de l'institutionnalisation du système, en précise au moins les contours (cf. Bougard 1995, p.180 et dans une moindre mesure, Kaiser 1986, p.95).

Un autre changement des dynamiques politiques se manifeste quant aux modalités de recrutement des *missi dominici* (Figure 5). L'ascension au pouvoir de Charles le Chauve a marqué une réduction considérable du cadre géographique de la provenance des agents. Il ne faut pas cependant perdre de vue les transformations profondes de la topographie du pouvoir qui affectaient l'Europe occidentale durant cette période. Les lots des héritiers du trône se resserraient comme une peau de chagrin sur la carte de l'ancien empire. Les limites de l'espace géographique dans lequel les souverains recrutaient leurs agents s'inscrivaient inévitablement dans cette conjoncture historique (cf. Bougard 1995, p.180). Charlemagne et Louis le Pieux, quant à eux, bien qu'ayant fait appel à des agents issus de l'espace impérial large encore unifié, ne semblent pas avoir suivi la même logique de recrutement. Si Charlemagne a souvent recouru à des agents qui sont en fonction dans les zones périphériques de son empire, son fils s'est appuyé davantage sur des *missi* issus plutôt du centre des territoires sous son contrôle. Là encore, il est imprudent de laisser de côté les mutations intervenues sous la gouvernance de Louis le Pieux. Ce dernier semblait vouloir, comme cela a été souligné maintes fois (Eckhardt 1956 ; Kikuchi 2012), inscrire le cadre géographique des activités missatiques dans le maillage des provinces ecclésiastiques reconstituées par tant d'efforts au siècle passé (Werner 1980, pp.197-198 ; cf. toutefois pour l'espace italien Bougard 1995, p.298). Seules les études futures pourront apporter une lumière suffisante pour éclaircir le dessin des circonscriptions missatiques et leur évolution dans le temps.

C'est également le successeur du premier empereur franc qui est apparu comme une passerelle importante dans la continuité du système missatique (Figure 6). Le fils de Charlemagne s'appuyait sur les agents déjà utilisés par ses prédécesseurs et « transmettait » les siens à ses héritiers. Tout bien considéré, la transition des *missi* entre les souverains s'inscrit parfaitement dans la logique de l'évolution de la situation politique et familiale du IX[e] siècle. Les camps adverses utilisaient des agents différents, les fils faisaient appel aux *missi* de leur père. Ce processus et ces agents ont été jusqu'à là peu connus par l'histoire de cette institution. À terme, l'étude d'un tel mécanisme peut être abordée de plusieurs façons. Avant tout, la question de l'importance de ce type d'agents se pose. Si, dans l'optique de l'analyse de réseau, ce travail prête à l'examen plus attentif de la position et du prestige de ces envoyés dans l'ensemble du réseau, pour un historien, c'est tout d'abord la question des figures influentes de la politique carolingienne qui est soulevée. Enfin, le processus de la transition des agents d'un règne à l'autre, d'une cour à l'autre, invite à réfléchir sur la notion d'héritage politique et sur la continuité des mêmes pratiques de gouvernance dans le temps (Airlie 1990 ; Airlie 2006).

Ces quelques premiers constats plaident en faveur du maintien des pratiques missatiques importantes durant le IX[e] siècle. Le nombre d'agents et de missions de chaque souverain a certes diminué, mais, en même temps, l'institution semblait évoluer vers une structure relativement plus complexe. Toujours en proie aux bouleversements qui secouent l'Europe occidentale, les mécanismes de recrutement et le cadre géographique changeaient et s'adaptaient avec le temps, mais le système est resté toutefois actif tout au long de la période étudiée.

Le deuxième volet des analyses s'est efforcé d'éclaircir la question du recrutement des *missi dominici* au sein de l'aristocratie locale. Les relations entre les lieux des missions et les endroits de fonctions et d'attaches personnelles ont été au centre des interrogations. La mesure des distances les plus courtes entre lesdites places a livré plusieurs indices fructueux et a confirmé les thèses historiographiques actuelles (Figures 7 et 9). Une grande partie des agents effectuaient leurs activités missatiques dans un rayon de 100-125 kilomètres de leurs lieux de pouvoir. L'absence d'un nombre important de données, notamment sur les liens de parenté, n'a pas permis toutefois de confirmer ce constat avec force. Les distances longues, quant à elles, dépassaient rarement 3000 kilomètres et semblaient être liées avant tout, comme cela a été montré, à l'ambiguïté du vocable *missi* employé tantôt pour les envoyés à l'intérieur de l'empire tantôt pour les délégués à l'étranger.

Ces distances ont évolué également dans le temps (Figures 8 et 10). Dès le deuxième tiers du IX[e] siècle, le recrutement des *missi dominici* se faisait de plus en plus au sein de l'aristocratie sur place. Hannig (1983), Werner (1980) et, en mesure moindre, Kaiser (1986) (cf. toutefois Bougard 1995, p.296) ont souligné les avantages de cette politique : les agents locaux disposaient déjà d'une autorité naturelle dans les zones de leurs affectations. Or, les raisons d'un tel changement de la logique de recrutement peuvent être multiples. Si Krause (1890) y voit un signe d'une dégradation inévitable de toute l'institution missatique au profit des pouvoirs régionaux, Werner (1980) préfère, avec plus de prudence, parler de mutations plus profondes, sociales et économiques, dans le système de la gouvernance carolingienne. À croire les analyses statistiques, la tendance de faire appel aux agents sur place a été pourtant présente dès la fin du VIII[e] siècle et n'a fait que se confirmer, à l'exception de règne de Charlemagne, durant le siècle suivant (Figure 7). Ce processus apparaît dès lors comme un mouvement complexe où le temps n'explique qu'une partie du phénomène.

Finalement, l'interprétation des résultats de l'analyse des distances doit être inscrite dans le contexte de l'histoire politique de la période étudiée. À partir des années 840, l'empire carolingien a été en proie à des partages successifs quand plusieurs nouveaux royaumes commençaient à émerger dans le paysage politique. La diminution des distances entre les lieux d'affectations et les lieux de fonctions d'un *missus* correspondait tout compte fait à un cadre plus restreint dans lequel opéraient de nouveaux rois. Le nombre d'agents ainsi que le rayon de leur recrutement et de leurs activités missatiques se sont réduits inévitablement (cf. Jégou 2010). D'ailleurs, le même constat a été déjà fait lors de l'analyse de l'espace géographique de provenance des agents des différents souverains (Figure 5).

L'évolution des pratiques est toujours accompagnée par des changements dans le vocabulaire employé. Dreillard (2001), Scior (2009) et Kikuchi (2013) l'ont déjà souligné : le terme *missi* pouvait être appliqué aussi bien aux agents en mission au coeur du royaume qu'à ceux envoyés sur des territoires plus éloignés. La diminution, avec le temps, des distances parcourues par les agents invite à s'interroger sur l'évolution terminologique opérant dans les sources quand le vocable *missi* a commencé à être réservé à un type plus spécifique d'agents.

En définitive, l'analyse des distances confirme les thèses historiographiques actuelles sur le recrutement, dans la majorité des cas, des *missi dominici* dans une forte proximité de leurs lieux d'affectations (Werner 1980 ; Eckhardt 1956 ; Kaiser 1986). Il n'est pas toutefois inutile de s'attarder sur la notion même de la *distance*, comprise et vécue différemment à l'époque médiévale (Power 1999 ; Jégou 2010 ; Gravel 2012, notamment pp.314-317). De fait, le calcul des distances les plus courtes est assez subjectif du point de vue historique. Une comparaison quantifiée des éloignements géographiques ne prend pas obligatoirement en compte les difficultés du paysage et la présence des voies de communication. Donc, retenons seulement ce que cette distance, qu'elle soit subjective ou non, peut représenter. Arn de Salzbourg a été certainement plus apte à avoir de l'influence en Bavière que Ingoubertus, comte de Rouen en marche d'Espagne, que ces endroits aient été reliés par des routes ou séparés par des montagnes (voir pour cet exemple *infra* Section II, cf. également Werner 1980, p.210).

Les futures pistes de recherches sont malgré tout nombreuses. Bougard (1995, p.188) a déjà signalé, dans le cadre italien notamment, un lien étroit des zones des activités missatiques avec les déplacements de la cour royale. Une analyse ultérieure de la topographie des missions sera plus que judicieuse. Les objectifs des missions, déjà mis à l'honneur par l'historiographie (Bougard 1995 ; Davis 2015 ; Depreux 2001 ; Kaiser 1986) et exclus, pour les raisons évoquées plus haut, du présent travail, méritent également une attention spéciale. Leur corrélation avec, par exemple, des fonctions exercées par les envoyés royaux apportera sans doute des témoignages nouveaux sur l'organisation de l'institution missatique. Quant à l'étude d'autres liens que ceux déjà observés à l'intérieur des réseaux des agents et des rois, un grand nombre de données incomplètes la rendent d'emblée périlleuse. Seul un dépouillement additionnel des sources sera en mesure de pallier ces lacunes.

Le manque de données a été du reste un des plus grands défis de cette recherche. L'objectif a été alors de montrer aussi bien les possibilités de l'interprétation que la fragilité des résultats issus de l'analyse d'un réseau politique délimitée par la documentation fragmentaire. De telles difficultés pourtant ne sont pas propres aux enquêtes historiques fondées sur des sources d'ordinaire partielles, mais communes à la majorité des disciplines (Beauguitte 2016). Un regard attentif sur toutes les données de l'histoire missatique recueillies à ce jour (Table 1) invite à une vigilance redoublée face à tout bilan, passé ou futur, qu'il est possible d'en tirer. Peut-on se prononcer de manière arrêtée sur les liens de parenté des *missi dominici*, alors que l'on ne dispose à cet égard d'informations que sur 14% d'entre eux ? Est-il judicieux de projeter ces conclusions sur l'ensemble de l'institution ? Les pratiques missatiques en Bavière étudiées par Hanning (1984b) ou en Italie traitées par Bougard (1995) ne sont que des cas spécifiques inscrits dans une conjoncture particulière. Ces interrogations, connues des historiens, peuvent trouver des réponses nouvelles dans le cadre méthodologique revisité : de l'échantillonnage déjà habituel aux techniques plus élaborées de traitement des données manquantes (voir par exemple Little et Rubin 2002). Si, pour l'heure, ce travail ne fait que le souligner, une fois de plus, les futures pistes de développement semblent se dessiner d'elles-mêmes.

Le changement d'approche nous a également invités à une vision plus large du système missatique. Si chaque cas est unique et si le contexte est indiscutablement important, embrasser toutes les données disponibles exige de centrer notre attention sur les modèles globaux qui se dégagent au fil des analyses. Le formalisme des techniques réticulaires appuyées par les démarches statistiques n'a pas effacé la singularité des faits, mais a apporté une vue du processus dans son ensemble et dans sa continuité. Les modèles de réseaux ainsi que les méthodes analytiques proposées apportent alors un regard nouveau, à travers l'exemple des agents du pouvoir central, sur l'un des aspects des dynamiques de gouvernance durant le premier Moyen Âge.

**Références**

**Sources citées**

Annales regni Francorum = *Annales regni Francorum inde a. 741 usque ad 829, qui dicuntur Annales Laurissenses maiores et Einhardi*, MGH, SS rer.Germ., 6, éd. Kurze, F., Pertzii G.H., Hannover, 1895, 2-178. http://www.mgh.de/dmgh/resolving/MGH_SS_rer._Germ._6_S._IV

Bernard, A., & Bruel, A. (Eds.). (1876). *Recueil des chartes de l'abbaye de Cluny* Vol. 1 (802-954). Paris. https://gallica.bnf.fr/ark:/12148/bpt6k28908j

Capit. I = *Capitularia regum Francorum I*, MGH, Leges, Boretius, A. (éd.), Hannover, 1883. http://www.mgh.de/dmgh/resolving/MGH_Capit._1_S._II

Capit. II = *Capitularia regum Francorum II*, MGH, Leges, Boretius, A., Krause, V. (éds.), Hannover, 1897. http://www.mgh.de/dmgh/resolving/MGH_Capit._2_S._II

Conc. 2.1 = *Concilia aevi Karolini (742-842), Teil 1 [742-817]*, MGH, Leges, Conc. 2.1, éd. Werminghoff, A., Hannovre, 1906. http://www.mgh.de/dmgh/resolving/MGH_Conc._2,1_S._II

Conc. 3 = *Die Konzilien der karolingischen Teilreiche 843-859*, MGH, Leges, Conc. 3, éd. Hartmann, A., Hannover, 1984. http://www.mgh.de/dmgh/resolving/MGH_Conc._3_S._II

Bitterauf, T., (Ed.). (1905). *Die Traditionen des Hochstifts Freising*, Munich, 1905. https://bildsuche.digitale-sammlungen.de/index.html?c=viewer&bandnummer=bsb00004628&pimage=138&v=100&nav=&l=en

DD LD = Die Urkunden Ludwigs des Deutsche, in *Ludwig der Deutsche, Karlmann und Ludwig der Jüngere*, MGH, Diplomata (DD), éd. Kehr. P., Berlin, 1934, 1-285. http://www.mgh.de/dmgh/resolving/MGH_DD_LdD_/_DD_Km_/_DD_LdJ_S._1

Theodulfi, Contra Iudices = *Versus Theodulfi episcopi contra Iudices* (Theodulfi Carmina, XXVIII), MGH, Antiquitates, Poetae latini Aevi Carolini I, éd. Dümmler E., Berlin, 1881, 493-517. http://www.mgh.de/dmgh/resolving/MGH_Poetae_1_S._493

**Bibliographie**

Airlie S. (1990). Bonds of Power and bonds of association in the court circle of Louis the Pious. In P. Godman & R. Collins (Eds.), *Charlemagne's Heir. New perspectives on the reign of Louis the Pious (814-840)*. Oxford : Clarendon Press, 191-205. http://opac.regesta-imperii.de/id/104719

Airlie S. (2006). The aristocracy in the service of the state in the Carolingian period. In S. Airlie, W. Pohl, & H. Reimitz (Eds.), *Staat im frühen Mittelalter*. Wien : Verlag der Österreichischen Akademie der Wissenschaften, 93-113. http://opac.regesta-imperii.de/id/1155837

Althoff G. (2004). *Family, Friends and Followers : Political and Social Bonds in Early Medieval Europe* : Cambridge University Press.

Bartholomew D., Steele F., Galbraith J., & Moustaki I. (2008). *Analysis of Multivariate Social Science Data* (2nd ed.) : CRC Press. https://www.crcpress.com/p/book/9781584889601

Beauguitte L. (2016). L'analyse de réseaux en sciences sociales et en histoire. Vocabulaire, principes et limites. In *Le réseau. Usages d'une notion polysémique en sciences humaines et sociales* : Presses Universitaires de Louvain, 9-24. https://pul.uclouvain.be/book/?GCOI=29303100292500

Borgolte M. (1986). *Die Grafen Alemanniens in merowingischer und karolingischer Zeit : eine Prosopographie*. Sigmaringen : J. Thorbecke. https://www.thorbecke.de/die-grafen-alemanniens-in-merowingischer-und-karolingischer-zeit-eine-prosopographie-p-610.html

Bougard F. (1995). *La justice dans le royaume d'Italie de la fin du VIII$^e$ siècle au début du XI$^e$ siècle*. Rome : École française de Rome. http://www.publications.efrome.it/opencms/opencms/_461e25d8-8c2e-11e0-9a66-000c291eeace.html

Bougard F., Bührer-Thierry G. v., & Le Jan R. (2013). Les élites du haut Moyen Âge : identités, stratégies, mobilité. *Annales. Histoire, Sciences Sociales, 4*, 1079-1112. https://www.cairn.info/revue-annales-2013-4-page-1079.htm

Bouveyron C., Jegou L., Jernite Y., Lamassé S., Latouche P., & Rivera P. (2014). The Random Subgraph Model for the Analysis of an Ecclesiastical Network in Merovingian Gaul. *The Annals of Applied Statistics, 8*(1), 377-405. https://www.jstor.org/stable/24521738

Brughmans T., Collar A., & Coward F. (Eds.). (2016). *The Connected Past. Challenges to Network Studies in Archaeology and History*. Oxford : Oxford University Press. https://global.oup.com/academic/product/209780198748519

Bühler A. (1986). Capitularia Relecta. Studien zur Entstehung und Überlieferung der Kapitularien Karls des Großen und Ludwigs des Frommen. *Archiv für Diplomatik, 32*, 305-502. http://opac.regesta-imperii.de/id/110439

Collar A., Brughmans T., Coward F., & Lemercier C. (2014). Analyser les réseaux du passé en archéologie et en histoire. *Les nouvelles de l'archéologie, 135*, 9-13. https://journals.openedition.org/nda/2300

Collar A., Coward F., Brughmans T., & Mills B. (2015). Networks in Archaeology : Phenomena, Abstraction, Representation. *Journal of Archaeological Method and Theory, 22*(1), 1-32. doi: 10.1007/s10816-014-9235-6

Collins T. (1950). Sur quelques vers de Theodulfe. *Revue bénédictine, 60*, 214-218. doi: 10.1484/J.RB.4.00027

Contamine P. (Ed.) (2002). *Le Moyen Age. Le roi, l'Église, les grands, le peuple 481-1514*. Paris. http://www.seuil.com/ouvrage/le-moyen-age-le-roi-l-eglise-les-grands-le-peuple-481-1514-philippe-contamine/9782757801864

Cox T., & Cox M. (2001). *Multidimensional Scaling* (2nd ed.) : Chapman & Hall/CRC. doi: 10.1007/978-3-540-33037-0_14

Crawley M. (2015). *Statistics. An Introduction using R* : Wiley. https://www.wiley.com/en-fr/Statistics:+An+Introduction+Using+R,+2nd+Edition-p-9781118941102

Davies R. (2003). The Medieval State : The Tyranny of a Concept? *Journal of Historical Sociology, 16*(2), 280-300. doi: 10.1111/1467-6443.00206

Davis J. R. (2015). *Charlemagne's Practice of Empire*. Cambridge : Cambridge University Press. doi: 10.1017/CBO9781139924726

de Clercq C. (1968). *Neuf capitulaires de Charlemagne concernant son œuvre réformatrice par les « Missi »*. Milano : A. Giuffre. https://ccfr.bnf.fr/portailccfr/ark:/06871/0011156338

De Jong M. (2009). *The penitential state : authority and atonement in the age of Louis the Pious, 814-840*. Cambridge : New York : Cambridge University Press.

Depreux P. (1997). *Prosopographie de l'entourage de Louis le Pieux (781-840)*. Thorbecke : Sigmaringen. https://www.perspectivia.net/publikationen/instrumenta/depreux_prosopographie

Depreux P. (2001). L'absence de jugement datant du règne de Louis le Pieux : l'expression d'un mode de gouvernement reposant plus systématiquement sur le recours aux missi. *Annales de Bretagne et des Pays de l'Ouest, 108*(1), 7-20.

Depreux P. (2011). *Dominus*. Marques de respect et expression des rapports hiérarchiques dans la désignation des personnes d'autorité. In F. Bougard, H.-W. Goetz, & R. Le Jan (Eds.), *Théorie et pratiques des élites au haut Moyen Age. Conception, perception et réalisation sociale* : Brepols, 187-221. doi: 10.1484/M.HAMA-EB.6.09070802050003050402020603

Depreux P., Bougard F., Le Jan R., & France. (2007). *Les élites et leurs espaces : mobilité, rayonnement, domination : du VI$^e$ au XI$^e$ siècle*. Turnhout (Belgique) : Brepols. doi: 10.1484/M.HAMA-EB.6.09070802050003050206010109

Dreillard R. (2001). La conclusion des traités à l'ère carolingienne. Une négociation « internationale » ? *Hypothèses, 4*(1), 171-179. doi: 10.3917/hyp.001.0171

Dumézil B. (2013). *Servir l'État barbare dans la Gaule franque. Du fonctionnariat antique à la noblesse médiévale, IV$^e$-IX$^e$ siècle*. Paris : Tallandier. https://catalogue.bnf.fr/ark:/12148/cb435792180

Ebling H. (1974). *Prosopographie der Amtsträger des Merowingerreiches. Von Chlothar II. (613) bis Karl Martell (741)*. München : Wilhelm Fink. https://www.perspectivia.net/receive/ploneimport_mods_00010192

Eckhardt W. A. (1956). Die Capitularia missorum specialia von 802. *Deutsches Archiv für Erforschung des Mittelalters, 12*, 498-516. http://www.digizeitschriften.de/dms/resolveppn/?PID=GDZPPN000351482

Faust K. (1997). Centrality in affiliation networks. *Social Networks, 19*(2), 157–191. doi: 10.1016/S0378-8733(96)00300-0

Fleckenstein J. (1991a). Karl der Große. In R. Auty & R.-H. Bautier (Eds.), *Lexikon des Mittelalters* (Vol. V). Munich : Artemis Verlag, col. 956-961. http://www.brepols.net/Pages/BrowseBySeries.aspx?TreeSeries=LEXMA-O

Fleckenstein J. (1991b). Ludwig (I.) der Fromme. In R. Auty & R.-H. Bautier (Eds.), *Lexikon des Mittelalters* (Vol. V). Munich : Artemis Verlag, col. 2171-2172. http://www.brepols.net/Pages/BrowseBySeries.aspx?TreeSeries=LEXMA-O

Fleckenstein J. (1993). Pippin III (der Jünger). In R. Auty & R.-H. Bautier (Eds.), *Lexikon des Mittelalters* (Vol. VI). Munich : Artemis Verlag, col. 2168-2170. http://www.brepols.net/Pages/BrowseBySeries.aspx?TreeSeries=LEXMA-O

Fried J. (1982). Der karolingische Herrschaftsverband im 9, Jh. zwischen « Kirche » und « Königshaus ». *Historische Zeitschrift, 235*(1), 1-43. doi: 10.1524/hzhz.1982.235.jg.1

Ganshof F.-L. (1958). *Recherches sur les capitulaires*. Paris : Sirey (impr. de Jouve). https://www.jstor.org/stable/43844577

Ganshof F.-L. (1965). Charlemagne et les institutions de la monarchie franque. In H. Beumann & W. Braunfels (Eds.), *Karl der Grosse. Lebenswerk und Nachleben* (Vol. 1). Düsseldorf, 349-393. http://opac.regesta-imperii.de/id/137675

Gauvard C. (2002). *La France au Moyen Age du V$^e$ au XV$^e$ siècle* (3 ed.). Paris : Presses universitaires de France. https://www.puf.com/content/La_France_au_Moyen_%C3%82ge_du_Ve_au_XVe_si%C3%A8cle

Goetz H.-W. (1987). Regnum : Zum politischen Denken in der Karolingerzeit. *Zeitschrift der Savigny-Stiftung für Rechtsgeschichte, Germanistische Abteilung, 104*, 110-189. http://opac.regesta-imperii.de/id/1309960

Goetz H.-W. (1991a). Lothar I. In R. Auty & R.-H. Bautier (Eds.), *Lexikon des Mittelalters* (Vol. V). Munich : Artemis Verlag, col. 2123-2124. http://www.brepols.net/Pages/BrowseBySeries.aspx?TreeSeries=LEXMA-O

Goetz H.-W. (1991b). Lothar II. In R. Auty & R.-H. Bautier (Eds.), *Lexikon des Mittelalters* (Vol. V). Munich : Artemis Verlag, col. 2124-2125. http://www.brepols.net/Pages/BrowseBySeries.aspx?TreeSeries=LEXMA-O

Goldberg E. J. (2006). *Struggle for Empire : kingship and conflict under Louis the German, 817-876*. Ithaca (N.Y.) : Cornell University Press. https://www.cornellpress.cornell.edu/book/9780801475290/struggle-for-empire/

Gravel M. (2007). Du rôle des missi impériaux dans la supervision de la vie chrétienne. Témoignage d'une collection de capitulaires du début du IX$^e$ siècle. *Memini. Travaux et documents, 11*, 99-130. https://journals.openedition.org/memini/128

Gravel M. (2012). *Distances, rencontres, communications : réaliser l'empire sous Charlemagne et Louis le Pieux*. Turnhout : Brepols. doi: 10.1484/M.HAMA-EB.4.00032

Grunin A. (2019). Le Moyen Âge, une époque sans État ? Construire le passé au présent. *Perspectives médiévales, 40*, 1-21. https://journals.openedition.org/peme/15221

Hamilton J. D. (1994). *Times Series Analysis* : Princeton University Press. https://press.princeton.edu/titles/5386.html

Hammond M., Jackson C., Bradley J., & Broun D. (2017). *Social Network Analysis and the People of Medieval Scotland (PoMS) 1093-1286 Database*. University of Glasgow : Centre for Scottish and Celtic Studies. https://www.poms.ac.uk/information/e-books/social-network-analysis-and-the-people-of-medieval-scotland-1093-1286-poms-database/

Hannig J. (1983). Pauperiores vassi de infra palatio ? Zur Entstehung der karolingischen Königsbotenorganisation. *Mitteilungen des Instituts für Österreichische Geschichtsforschung, 91*, 309-374. doi: 10.7767/miog.1983.91.34.309

Hannig J. (1984a). Zentrale Kontrolle und regionale Machtbalance : Beobachtungen zum System der karolingischen Königsboten am Beispiel des Mittelrheingebietes. *Archiv für Kulturgeschichte, 66*(1), 1-46. doi: 10.7788/akg-1984-0102

Hannig J. (1984b). Zur Funktion der karolingischen « missi dominici » in Bayern und in den südöstlichen Grenzgebieten. *Zeitschrift der Savigny-Stiftung für Rechtsgeschichte, Germanistische Abteilung, 101*, 256-300. doi: 10.7767/zrgga.1984.101.1.256

Hlawitschka E. (1960). *Franken, Alemannen, Bayern und Burgunder in Oberitalien (774-962). Zum Verständnis der Fränkischen Königsherrschaft in Italien* (Vol. 8). Freiburg im Breisgau : Eberhard Albert Verlag. http://www.manfred-hiebl.de/genealogie-mittelalter/hlawitschka_franken_alemannen/index.htm

Hudson P., & Ishizu M. (2016). *History by Numbers. An Introduction to Quantitative Approaches* : Bloomsbury. https://www.bloomsbury.com/uk/history-by-numbers-9781474294157/

Jarnut J. (1991). Karlmann, Frankenreich. In R. Auty & R.-H. Bautier (Eds.), *Lexikon des Mittelalters* (Vol. V). Munich : Artemis Verlag, col. 996. http://www.brepols.net/Pages/BrowseBySeries.aspx?TreeSeries=LEXMA-O

Jégou L. (2010). Les déplacements des 'missi dominici' dans l'Empire carolingien (fin VIII$^e$ - fin IX$^e$ siècle). In *Des sociétés en mouvement. Migrations et mobilité au Moyen Âge.* Paris : Publications de la Sorbonne, 223-235. https://books.openedition.org/psorbonne/10452

Kadushin C. (2012). *Understanding Social Networks : Theories, Concepts, and Findings* : Oxford University Press. https://global.oup.com/academic/product/understanding-social-networks-9780195379471

Kaiser R. (1986). Les évêques de Langres dans leur fonction de *Missi dominici*. In *Aux origines d'une seigneurie ecclésiastique. Langres et ses évêques VIII$^e$-XI$^e$ siècles.* Langres, 91-113. https://www.persee.fr/doc/bec_0373-6237_1988_num_146_2_450522_t1_0439_0000_5

Kikuchi S. (2012). *Carolingian capitularies as texts : significance of texts in the government of the Frankish kingdom especially under Charlemagne.* Configuration du texte en histoire. Global COE Program International Conference Series, Nagoya University.

Kikuchi S. (2013). *Untersuchungen zu den Missi dominici. Herrschaft, Delegation und Kommunikation in der Karolingerzeit.* (Doktorgrades der Philosophie), Ludwig-Maximilians-Universität München, http://opac.regesta-imperii.de/id/2166204

Kolenikov S., Steinley D., & Thombs L. (Eds.). (2010). *Statistics in the Social Sciences. Current methodological developments* : Wiley. https://www.wiley.com/en-us/Statistics+in+the+Social+Sciences%3A+Current+Methodological+Developments-p-9780470148747

Krause V. (1890). Geschichte des Institutes der missi dominici. *Mitteilungen des Instituts für Österreichische Geschichtsforschung, 11*, 193–300. doi: 10.7767/miog.1890.11.jg.193

Lemercier C. (2005). Analyse de réseaux et histoire. *Revue d'histoire moderne et contemporaine, 2*(52-2), 88-112. doi: 10.3917/rhmc.522.0088

Lemercier C., & Zalc C. (2019). *Quantitative Methods in the Humanities* : University of Virginia Press. https://www.upress.virginia.edu/title/5168

Little R., & Rubin D. (2002). *Statistical analysis with missing data* : Wiley. doi: 10.1002/9781119013563

Magnou-Nortier É. (1994). La mission financière de Théodulf en Gaule méridionale d'après les « Contra iudices ». In *Papauté, monachisme et diéories politiques. Mélanges Marcel Pacaut*, 89-110. http://presses.univ-lyon2.fr/produit.php?id_produit=241

Manaresi C. (Ed.) (1955). *I placiti del « Regnum Italiae »* (Vol. I, a. 776-945). Rome.

McCormick M. (2001). *Origins of the European economy : communications and commerce, A.D. 300-900* : Cambridge University Press. doi: 10.1017/CBO9781107050693

McCormick M. (2011). *Charlemagne's survey of the Holy Land. Wealth, personnel, and buildings of a Mediterranean church between Antiquity and the Middle Ages. With a critical edition and translation of the original text* : Dumbarton Oaks. https://www.doaks.org/research/publications/books/charlemagne2019s-survey-of-the-holy-land

McKitterick R. (2008). *Charlemagne : the formation of a European identity.* Cambridge, New York : Cambridge University Press.

McKitterick R. (2009). Charlemagne's missi and their books. In S. Baxter, C. Karkov, J. Nelson, & D. Pelteret (Eds.), *Early medieval studies in memory of Patrick Wormald* : Aldershot, 253-268. doi: 10.4324/9781315257259

Nelson J. (2000). Messagers et intermédiaires en Occident et au-delà à l'époque carolingienne. In A. Dierkens & J.-M. Sansterre (Eds.), *Voyages et voyageurs à Byzance et en Occident du VI$^e$ au XI$^e$ siècle.* Genève, 397-413. http://opac.regesta-imperii.de/id/101685

Newman M. (2010). *Networks. An Introduction* : Oxford University Press. doi: 10.1093/acprof:oso/9780199206650.001.0001

Poly J.-P. (1976). *La Provence et la société féodale 879-1166.* Paris. https://gallica.bnf.fr/ark:/12148/bpt6k33313593.texteImage

Power D. (1999). Frontiers : Terms, Concepts, and the Historians of Medieval and Early Modern Europe. In D. Power & N. Standen (Eds.), *Frontiers in Question. Eurasian Borderlands, 700–1700* : Palgrave Macmillan, 1-31. http://opac.regesta-imperii.de/id/192461

Reynolds S. (1997). The historiography of the medieval state. In M. Bentley (Ed.), *Companion to historiography.* London : Routledge, 109-130. https://www.crcpress.com/Companion-to-Historiography/Bentley/p/book/9780415285575

Riché P. (1997). *Les Carolingiens : Une famille qui fit l'Europe.* Paris. https://www.persee.fr/doc/rhr_0035-1423_1985_num_202_3_2741

Rio A., Bradley J., Nelson J. L., Roberts E., Stone R., Pasin M., de Angelis G. (2014). The Making of Charlemagne's Europe (768-814). from King's College London http://www.charlemagneseurope.ac.uk

Rosé I. (2011). Reconstitution, représentation graphique et analyse des réseaux de pouvoir au haut Moyen Âge. Approche des pratiques sociales de l'aristocratie, à partir de l'exemple d'Odon de Cluny († 942). *Redes, Redes sociales e Historia, 21*, 199-272. https://www.raco.cat/index.php/Redes/article/view/249782

Schneider R. (1997). Der rex Romanorum als gubernator oder administrator imperii. *Zeitschrift der Savigny-Stiftung für Rechtsgeschichte, Germanistische Abteilung, 114*, 296-317.

Schneidmüller B. (1991a). Karl (II) der Kahle. In R. Auty & R.-H. Bautier (Eds.), *Lexikon des Mittelalters* (Vol. V). Munich : Artemis Verlag, col. 967-968. http://www.brepols.net/Pages/BrowseBySeries.aspx?TreeSeries=LEXMA-O

Schneidmüller B. (1991b). Ludwig (II.) 'der Stammler'. In R. Auty & R.-H. Bautier (Eds.), *Lexikon des Mittelalters* (Vol. V). Munich : Artemis Verlag, col. 2175-2176. http://www.brepols.net/Pages/BrowseBySeries.aspx?TreeSeries=LEXMA-O

Schneidmüller B. (1991c). Karl (III) der Dicke. In R. Auty & R.-H. Bautier (Eds.), *Lexikon des Mittelalters* (Vol. V). Munich : Artemis Verlag, col. 968-969. http://www.brepols.net/Pages/BrowseBySeries.aspx?TreeSeries=LEXMA-O

Schneidmüller B. (1993a). Pippin (Karlmann). In R. Auty & R.-H. Bautier (Eds.), *Lexikon des Mittelalters* (Vol. VI). Munich : Artemis Verlag, col. 2171. http://www.brepols.net/Pages/BrowseBySeries.aspx?TreeSeries=LEXMA-O

Schneidmüller B. (1993b). Pippin I. In R. Auty & R.-H. Bautier (Eds.), *Lexikon des Mittelalters* (Vol. VI). Munich : Artemis Verlag, col. 2170. http://www.brepols.net/Pages/BrowseBySeries.aspx?TreeSeries=LEXMA-O

Scior V. (2009). Bemerkungen zum frühmittelalterlichen Boten- und Gesandtschaftswesen. In W. Pohl & V. Wieser (Eds.), *Der frühmittelalterliche Staat – europäische Perspektiven*. Wien : Verlag der Österreichische Akademie der Wissenschaften, 315-329. http://www.austriaca.at/0xc1aa500e_0x0023f6a1.pdf

Scott J. (2000). *Social Network Analysis. A Handbook* (2nd ed.) : SAGE. https://uk.sagepub.com/en-gb/eur/social-network-analysis/book249668

Simonoff J. (1996). *Smoothing methods in Statistics* : Springer. doi: 10.1007/978-1-4612-4026-6

Sot M. (2007). Références et modèles romains dans l'Europe carolingienne. Une approche iconographique du prince. In J.-P. Genet (Ed.), *Rome et l'Etat moderne européen*. Rome : École française de Rome, 15-42. http://www.publications.efrome.it/opencms/opencms/rome_et_l'%C3%A8tat_moderne_europ%C3%A8en_43e7e5a6-8c2e-11e0-9a66-000c291eeace.html

Störmer W. (1991a). Ludwig II der Deutsche. In R. Auty & R.-H. Bautier (Eds.), *Lexikon des Mittelalters* (Vol. V). Munich : Artemis Verlag, col. 2172-2174. http://www.brepols.net/Pages/BrowseBySeries.aspx?TreeSeries=LEXMA-O

Störmer W. (1991b). Karlmann, Ostfrankenreich. In R. Auty & R.-H. Bautier (Eds.), *Lexikon des Mittelalters* (Vol. V). Munich : Artemis Verlag, col. 996. http://www.brepols.net/Pages/BrowseBySeries.aspx?TreeSeries=LEXMA-O

Tessier G. (1967). *Charlemagne*. Paris : Albin Michel.

Thompson J. W. (1903). The decline of the missi Dominici in Frankish Gaul. *The decennial publications, 4*, 291-310. https://archive.org/details/declineofmissido00thomrich/page/n6

Thompson S. (2012). *Sampling* (3 ed.) : Wiley. doi: 10.1002/9781118162934

Tignolet C. (2011). Mobiliser des soutiens. L'échec de Théodulfe d'Orléans (vers 750/760 – 820/821). *Hypothèses, 14*(1), 259-268. doi: 10.3917/hyp.101.0259

Vauchez A. (1997). *Dictionnaire encyclopédique du Moyen âge avec la collab. de Catherine Vincent*. Paris : Éd. du Cerf.

Wasserman S., & Faust K. (1994). *Social Network Analysis : methods and applications* : Cambridge University Press. https://www.cambridge.org/zw/academic/subjects/sociology/sociology-general-interest/social-network-analysis-methods-and-applications

Werner K. F. (1980). Missus - marchio – comes : entre l'administration centrale et l'administration locale de l'Empire carolingien. In W. Paravicini (Ed.), *Histoire comparée de l`administration*. Zürich : Artemis Verl., 191-239. https://www.perspectivia.net/publikationen/bdf/paravicini-werner_administration/werner_missus

Werner K. F. (1981). *La genèse des duchés en France et en Allemagne*. Nascita dell'Europa ed Europa carolingia : un'equazione da verificare, Spolète. http://opac.regesta-imperii.de/id/216670

Werner K. F. (1992). L'historien et la notion d'État. *Comptes-rendus des séances de l'Académie des inscriptions et belles-lettres, 136*(4), 709-721. https://www.persee.fr/doc/crai_0065-0536_1992_num_136_4_15149

White D. R., & Borgatti S. P. (1994). Betweeness centrality measures for directed graphs. *Social Networks, 16*, 335-346. doi: 10.1016/0378-8733(94)90015-9

Wilson T. P. (1982). Relational networks : an extension of sociometric concepts. *Social Networks, 4*, 105-116. doi: 10.1016/0378-8733(82)90028-4

Zielinski H. (1991). Ludwig II. In R. Auty & R.-H. Bautier (Eds.), *Lexikon des Mittelalters* (Vol. V). Munich : Artemis Verlag, col. 2177. http://www.brepols.net/Pages/BrowseBySeries.aspx?TreeSeries=LEXMA-O

## A  Remerciements